\documentclass[preprint]{revtex4}

\usepackage{scrpage2}			
\usepackage{graphicx}			
\usepackage{amsmath}			
\usepackage{amsfonts}
\usepackage{amssymb}
\usepackage{amsthm}
\usepackage{xcolor}

\newcommand{\dslash}{\not{\hbox{\kern-2pt $\partial$}}}
\newcommand{\bq}{\begin{equation}} 
\newcommand{\eq}{\end{equation}}
\newcommand{\bqa}{\begin{eqnarray}} 
\newcommand{\eqa}{\end{eqnarray}}
\newcommand{\nn}{\nonumber \\}

\newcommand{\msk}{\medskip}

\newcommand{\beq}{\begin{equation}}
\newcommand{\eeq}{\end{equation}}

\renewcommand{\vec}[1]{{\mathbf{#1}}}

\def\be{\begin{eqnarray}}
\def\ee{\end{eqnarray}}

\def\part{\partial}

\def\Dslash{\,\,{\raise.15ex\hbox{/}\mkern-13mu D}}
\def\Dbarslash{\,\,{\raise.15ex\hbox{/}\mkern-12mu {\bar D}}}
\def\delslash{\,\,{\raise.15ex\hbox{/}\mkern-10mu \partial}}
\def\delbarslash{\,\,{\raise.15ex\hbox{/}\mkern-9mu {\bar\partial}}}
\def\pslash{\,\,{\raise.15ex\hbox{/}\mkern-11mu p}}
\def\qslash{\,\,{\raise.15ex\hbox{/}\mkern-9mu q}}
\def\kslash{\,\,{\raise.15ex\hbox{/}\mkern-11mu k}}
\def\eslash{\,\,{\raise.15ex\hbox{/}\mkern-9mu \epsilon}}

\newcommand{\slsh}[1]{\,\,{\raise.15ex\hbox{/}\mkern-12mu {#1}}}

\begin{document}



\title{
Perturbative non-Fermi liquids from dimensional regularization 
}

\author{Denis Dalidovich$^{1}$ and Sung-Sik Lee$^{1,2}$\\
{\normalsize{$^1$Perimeter Institute for Theoretical Physics,}}\\
{\normalsize{31 Caroline St. N., Waterloo ON N2L 2Y5, Canada}}
\vspace{0.2cm}\\
{\normalsize{$^2$Department of Physics $\&$ Astronomy, McMaster University,}}\\
{\normalsize{1280 Main St. W., Hamilton ON L8S 4M1, Canada}}
}

\date{\today}

\begin{abstract}

We devise a dimensional regularization scheme for 
quantum field theories with Fermi surface
to study scaling behaviour of non-Fermi liquid states 
in a controlled approximation.
Starting from a Fermi surface in two space dimensions,
the co-dimension of Fermi surface is extended to
a general value while the dimension of Fermi surface is fixed.
When Fermi surface is coupled with a critical boson centred at zero momentum, 
the interaction becomes marginal at a
critical space dimension $d_c=5/2$.
A deviation from the critical dimension 
is used as a small parameter for a systematic expansion.
We apply this method to the theory where two patches of Fermi surface
is coupled with a critical boson, 
and show that the Ising-nematic critical point is described 
by a stable non-Fermi liquid state 
slightly below the critical dimension.
Critical exponents are computed up to the two-loop order.

\end{abstract}

\maketitle

\section{Introduction}

It is of central importance in condensed matter physics
to understand universal properties of phases
using low energy effective theories.
In critical states of quantum matter,  effective theories
take the form of quantum field theories 
which describe low energy degrees of freedom and their interactions.
Although the most generic critical state of electrons in solids is metal,
quantum field theories of metals are less well understood
compared to relativistic field theories
due to low symmetry and 
extensive gapless modes 
that need to be kept in low energy theories.

In Fermi liquid metals\cite{LANDAU}, quasiparticles provide a single-particle basis 
in which the low energy field theories can be diagonalized\cite{POLCHINSKI_FL,SHANKAR}.
In non-Fermi liquid states, there exist no such single-particle basis,
and the low energy physics is described by genuine interacting quantum 
field theories. 
Non-Fermi liquid states can arise 
when Fermi surface is coupled with a gapless boson
in many different physical contexts.
A boson can be made gapless either 
by fine tuning of microscopic parameters 
entailing non-Fermi liquid state at a quantum critical point,
or as a result of dynamical tuning, 
which gives rise to non-Fermi liquid phases 
within an extended region in the parameter space.
The examples for the former case include
heavy fermion compounds near magnetic quantum critical points\cite{LOHNEYSEN,COLEMAN},
quantum critical point for Mott transitions\cite{Senthil_Mott,Podolsky},
and the nematic quantum critical point \cite{ogankivfr,metzner,delanna,kee,lawler,rech,wolfle,maslov,quintanilla,yamase1,yamase2,halboth,jakub,zacharias,kim,huh}. 
The $\nu=1/2$ quantum Hall state\cite{HALPERIN} 
and Bose metals 
which support fractionalized fermionic excitations 
along with an emergent gauge field\cite{MOTRUNICH,LEE_U1,PALEE,MotrunichFisher}
are among the examples for the latter case.

From earlier works\cite{holstein,reizer,lee89,leenag},
it was pointed out that the low energy properties of Fermi surface
can be qualitatively modified by the coupling with gapless boson.
In three space dimensions,
logarithmic corrections arise
due to the Yugawa coupling\cite{holstein,reizer,Kachru}.
In two space dimensions,
theories of non-Fermi liquids
flow to strongly interacting
fixed points at low energies.
For chiral non-Fermi liquid states\cite{SSLee},
where only one patch of Fermi surface is coupled with a critical boson,
exact dynamical information can be extracted
thanks to the chiral nature of the theory\cite{shouvik}.
It is much harder to understand 
non-chiral theories which
include two-patches of Fermi surface
with opposite Fermi velocities. 
One strategy to make a progress in non-chiral theories is 
to deform the original theory 
into a perturbatively solvable regime in a continuous way.

There can be different and complimentary ways 
to obtain perturbative non-Fermi liquid states.
One attempt is to introduce a large number of flavors\cite{ALTSHULER,polchinski,ybkim}.
This is arguably the most natural extension
in the sense
that extra flavors do not introduce 
qualitatively new element to the theory
except that the flavor symmetry group is enlarged.
However,  it turns out that even the infinite flavor limit is not 
described by a mean-field theory
due to a large residual quantum fluctuations of Fermi surface\cite{SSLee,metlsach1}.
One way to achieve a controlled expansion 
is to deform the dynamics of the theory,
for example the dispersion of a critical boson,
to suppress quantum fluctuations at low energies\cite{nayak,mross}.
One can also try to access
two-dimensional non-Fermi liquid states
by increasing the number of one-dimensional chains,
where bosonization provides a controlled analytical tool\cite{Jiang}.
Finally, one can modify the dimension of spacetime continuously
to gain a controlled access to non-Fermi liquid states.
In doing so, one can extend either 
the dimension of Fermi surface\cite{Chakravarty}
 or the co-dimension\cite{senshank}.
In this paper, we devise a dimensional regularization scheme
where the co-dimension of Fermi surface is extended to
obtain a perturbative non-Fermi liquid state 
which describes two patches of Fermi surface 
coupled with a critical boson.
This scheme has an advantage that Fermi surface remains one-dimensional
and has only one tangent vector.
Because fermions in one region of the momentum space 
near the Fermi surface
are primarily coupled with the boson whose momentum
is tangential to the Fermi surface, 
fermions in different momentum patches 
(except for the ones in the exact opposite direction) 
are decoupled from each other in the low energy limit.
Because of this, one can focus on 
local patches in the momentum space,
which allows one to develop a systematic 
field theoretic renormalization group scheme.
If the dimension of Fermi surface is extended,
each patch has more than one tangent vectors.
Since one can not ignore couplings between different patches, 
the whole Fermi surface has to be included in the low energy theory.
Recently, a non-Fermi liquid state was studied
through a Wilsonian renormalization group scheme 
in $d=3-\epsilon$ space dimensions
with co-dimension of Fermi surface fixed to be one\cite{Fit}.

The paper is organized as follows.
In Sec. II, we introduce a $(2+1)$-dimensional theory 
which describes two patches of Fermi surface coupled
with a critical boson.
Depending on the way the boson is coupled with the patches,
the theory describes either the Ising-nematic critical point
or the quantum electrodynamics with a finite density.
In this paper, we will focus on the Ising-nematic theory.
We then generalize the theory to a $(d+1)$-dimensional theory 
with general space dimension $d$.
For this, we first combine particle in one patch
and hole in the opposite patch to construct 
a spinor with two components.
In this representation, fermionic excitations near the Fermi surface 
in the original $(2+1)$-dimensional theory
can be formally viewed as $(1+1)$-dimensional
Dirac fermions with a continuous flavor that 
corresponds to the momentum along the Fermi surface.
Then the Dirac fermion 
is extended to general dimensions, 
where the energy of fermion disperses linearly 
away from the gapless point in $(d-1)$ directions. 
Physically, this describes a one-dimensional Fermi line
embedded in $d$-dimensional momentum space.
In $d=3$ with $SU(2)$ flavor (spin) group, 
this theory describes a $p$-wave spin-triplet
superconducting state which supports a line node.
The Yugawa coupling between fermion and boson
becomes marginal at the critical dimension, $d_c=5/2$, 
and non-Fermi liquid states arise in $d<5/2$.
Using $\epsilon = 5/2-d$ as  an expansion parameter,
one can access the non-Fermi liquid state perturbatively.
The following section is devoted to the RG analysis of the theory
based on the dimensional regularization scheme.
In Sec. III. A, the minimal local action is constructed.
In Sec. III. B, the symmetry of the regularized theory is discussed.
Because the extension to higher dimensions involves 
turning on flavor non-singlet superconducting order parameter, 
the regularized theory breaks the charge conservation and some of the flavor symmetry.
In Sec. III. C, the renormalization group equation is derived using the
minimal subtraction scheme.
In Sec. III. D, we demonstrate that the expansion is controlled
in the small $\epsilon$ limit with fixed $N$ 
where $N$ is the number of fermion flavour.
In Sec. III. E, we summarize the computation of the counter terms up to the two-loop level. 
Some three-loop results are also included.
Based on the results, the dynamical critical exponents and the anomalous  dimensions are computed in Sec. III. F. 
While the interaction modifies 
the dynamics of  fermion in a non-trivial way, 
the boson does not receive a non-trivial quantum correction
up to the three-loop diagrams that we checked.
In Sec. IV, the scaling forms of the thermodynamic quantities 
and $2k_F$ scattering processes are obtained.
We finish with a summary and some outlook in Sec. V.
Details on the computation of Feynman diagrams 
can be found in the appendices.

\section{Model}

\begin{figure}
\begin{center}
        \includegraphics[height=5cm,width=6cm]{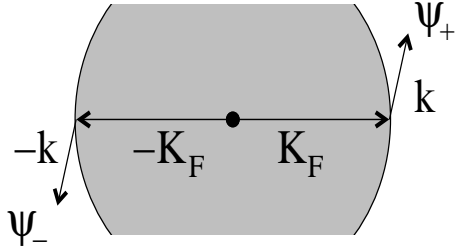} 
\end{center}
\caption{The right-moving and left-moving modes near the Fermi surface
can be combined into a two-component Dirac fermion.}
\label{fig:FS_two_patch}
\end{figure}

We consider a theory where two patches of Fermi surface
are coupled with one critical boson in $(2+1)$-dimensions,
\bqa
S & = &  \sum_{s=\pm} \sum_{j=1}^N \int \frac{d^3k}{(2\pi)^{3}}  
\psi_{s,j}^\dagger (k)
\Bigl[ 
 i k_0   +  s  k_1 +  k_2^2  \Bigr] \psi_{s,j}(k) \nonumber \\
 &+& \frac{1}{2} \int  \frac{d^3k}{(2\pi)^{3}} 
 \left[ k_0^2 + k_1^2 +  k_2^2 \right] \phi(-k) \phi(k) \nonumber \\
 &+&  \frac{e}{\sqrt{N}} \sum_{s=\pm} \sum_{j=1}^N  
\int \frac{d^3k d^3q}{(2\pi)^{6}}  ~
\lambda_s ~ \phi(q) ~  \psi^\dagger_{s,j}(k+q) \psi_{s,j}(k).
\label{act0}
\eqa
Here $\psi_{+,j}$ ($\psi_{-,j}$)
is the right (left) moving fermion with flavor $j=1,2,..,N$
whose Fermi velocity along the $k_1$ direction is positive (negative). 
The momenta are rescaled in such a way 
that the absolute value of Fermi velocity
and curvature of the Fermi surface are equal to one. 
$\phi$ is a real critical boson,
and $e$ is the fermion-boson coupling.
Although the velocity of boson is in general different from that of fermion,
the dynamics of boson is dominated by 
particle-hole excitations of fermion at low energies. 
As a result, the bare velocity of boson,
which is also set to be one in the action,  
does not matter for the low energy effective theory.
$\lambda_s$ controls the way the fermions are coupled with the boson.
The case with $\lambda_+ = \lambda_- = 1$ describes 
the Ising-nematic critical point.
The coupling with $\lambda_+ = -\lambda_- = 1$ describes
the quantum electrodynamics at a finite density,
where $\phi$ corresponds 
to the transverse component of the U(1) gauge field.
The action in Eq. (\ref{act0}) 
admits a self-contained 
renormalization group analysis\cite{metlsach1}.

In the following, we will focus on the Ising-nematic critical point.
Quantum phase transitions to nematic states with 
broken point group symmetry\cite{pomeran} have been observed 
in cuprate superconductors\cite{ando,hinkov,kohsaka,taill}, 
ruthenades\cite{borzi}, and pnictides\cite{fang1,xu1,chuang1,chu1}. 
The Ising-nematic order parameter is 
represented by a real scalar boson 
which undergoes 
strong quantum fluctuations 
at the quantum critical point.

Because the energy of fermion disperses only in one direction near the Fermi surface,
the $(2+1)$-dimensional fermion can be viewed as $(1+1)$-dimensional Dirac fermions,
where the momentum along the Fermi surface is interpreted as a continuous flavor.
To make this more precise, the right and left moving fermions are combined into 
one spinor (see Fig. \ref{fig:FS_two_patch}), 
\bqa
\Psi_j(k) = \left( 
\begin{array}{c}
\psi_{+,j}(k) \\
\psi_{-,j}^\dagger(-k)
\end{array}
\right).
\eqa
In this representation, the action in Eq. (\ref{act0}) becomes
\bqa
S & = &  \sum_{j} \int \frac{d^3k}{(2\pi)^{3}}  
\bar \Psi_j(k)
\Bigl[ 
 i k_0 \gamma_0  + i (   k_1 +  k_2^2 )  \gamma_1 \Bigr] 
\Psi_{j}(k) \nonumber \\
 &+& \frac{1}{2} \int  \frac{d^3k}{(2\pi)^{3}} 
 \left[ k_0^2 + k_1^2 +  k_2^2 \right] \phi(-k) \phi(k) \nonumber \\
 &+&     \frac{e}{\sqrt{N}} \sum_{j}  \int \frac{d^3k dq}{(2\pi)^{6}}  
\phi(q) \bar \Psi_{j}(k+q) W \Psi_{j}(k),
\label{act1}
\eqa
where $\gamma_0 = \sigma_y$, $\gamma_1 = \sigma_x$ 
are the gamma matrices for the two component spinor,
and $\bar \Psi \equiv \Psi^\dagger \gamma_0$.
$W = i \gamma_1$ ($W = \gamma_0$) 
for the Ising-nematic system (quantum electrodynamics). 
For the rest of the paper, we will focus on the Ising-nematic case.
The fermionic kinetic term is indeed identical 
to that of the $(1+1)$-dimensional Dirac fermion 
where the location of Dirac point
depends on $k_2$.

\begin{figure}
\begin{center}
        \includegraphics[height=6cm,width=14cm]{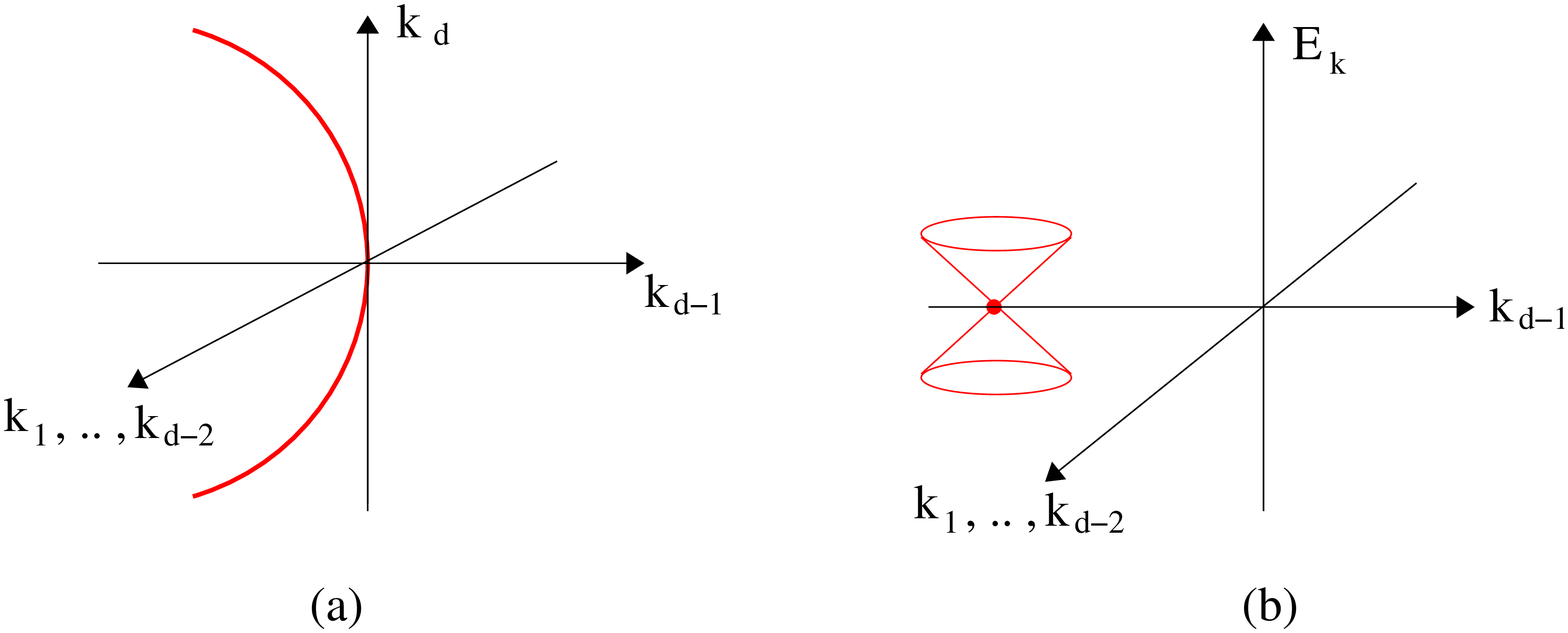} 
\end{center}
\caption{
(a) The one-dimensional Fermi surface embedded in the $d$-dimensional momentum space.
(b) For each $k_d$, there is a Fermi point at 
$(k_1, k_2, ..., k_{d-2}, k_{d-1}) = (0,0,.., 0, - \sqrt{d-1} k_d^2)$
which is denoted as a (red) dot.
Around the Fermi point, the energy disperses linearly
like a two-component Dirac fermion in 
$(d-1)$-dimensional space.
}
\label{fig:FS_general}
\end{figure}

Now we promote the theory to 
general dimensions.
The action that describes Fermi surface 
with a general co-dimension is written as
\begin{eqnarray}\label{act4}
S & = &  \sum_{j} \int \frac{d^{d+1}k}{(2\pi)^{d+1}} \bar \Psi_j(k)
\Bigl[ 
i \vec \Gamma \cdot \vec K + i \gamma_{d-1} \delta_k \Bigr] \Psi_{j}(k) \nonumber\\
&+&
\frac{1}{2} \int  \frac{d^{d+1}k}{(2\pi)^{d+1}}
 \left[ |\vec K|^2 + k_{d-1}^2 +  k_d^2 \right] \phi(-k) \phi(k) \nonumber \\
 &+&     \frac{ie}{\sqrt{N}}\sqrt{d-1} \sum_{j}  
\int \frac{d^{d+1}k d^{d+1}q}{(2\pi)^{2d+2}}  
\phi(q) \bar \Psi_{j}(k+q) \gamma_{d-1} \Psi_{j}(k).
\end{eqnarray}
Here $\vec K \equiv (k_0, k_1,\ldots, k_{d-2})$
represents frequency and $(d-2)$ components of  
the full $(d+1)$-dimensional energy-momentum vector,
$(k_0, k_1, ..., k_d)$.
$k_1, .., k_{d-2}$ are the newly added directions
which are transverse to the Fermi surface.
$\delta_k =  k_{d-1}+ \sqrt{d-1} k_d^2$
is the energy dispersion of the fermion within 
the original two-dimensional momentum space.
The gamma matrices associated with $\vec K$ 
are written as
$\vec \Gamma \equiv (\gamma_0, \gamma_1,\ldots, \gamma_{d-2})$.
Since the actual space dimension of 
interest lies between $2$ and $3$,
the number of spinor components is fixed to be two. 
We will use the representation
where $\gamma_0 = \sigma_y$ and $\gamma_{d-1} = \sigma_x$ are 
fixed in general dimensions.

The spinor has the energy dispersion with two bands,
\bqa
E_k = \pm \sqrt{ \sum_{i=1}^{(d-2)} k_i^2 + \delta_k^2  }.
\eqa
The energy vanishes if
\bqa\label{fersur2}
k_i &= &0,\qquad \mbox{for}\,\,\, i=1,\ldots,d-2\, \nn
{ k}_{d-1} & =& - \sqrt{d-1} k_d^2.
\eqa
Therefore the Fermi surface is a one-dimensional manifold 
embedded in the $d$-dimensional momentum space
as is shown in Fig. \ref{fig:FS_general}.

\begin{figure}
\begin{center}
        \includegraphics[height=6cm,width=8cm]{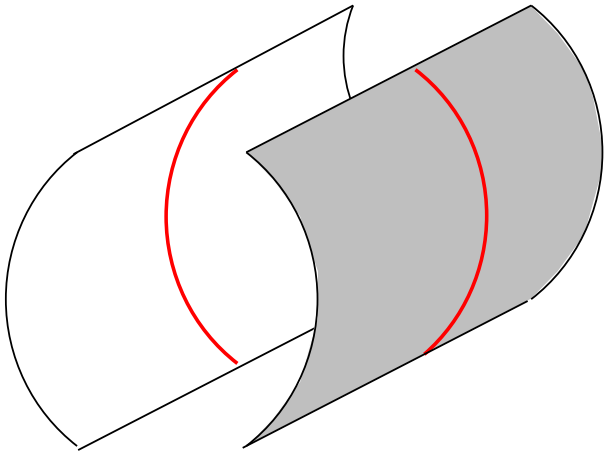} 
\end{center}
\caption{
Fermi lines in three dimensional momentum space can be  
obtained by turning on 
$p$-wave superconducting order parameter that
gaps out the cylindrical Fermi surface except 
for the line node denoted by the thick (red) line.
}
\label{fig:FS_3d}
\end{figure}

Before we delve into the RG analysis in an abstract dimension $d$,
we consider a concrete physical realization 
of the theory in $d=3$ with $N=2$, 
where two flavors represent spin $1/2$ degree of freedom.
In the basis 
where $\gamma_0 = \sigma_y$, $\gamma_1 = \sigma_z$, $\gamma_2 = \sigma_x$,
the quadratic action for the fermions becomes
\bqa
S & = &  
\int \frac{d^4k}{(2\pi)^{3}}  
\Bigg\{
\sum_{s=\pm} 
\sum_{j=\uparrow, \downarrow}
\psi_{s,j}^\dagger (k)
\left(  i k_0   +  s  k_2 + \sqrt{2} k_3^2  \right) \psi_{s,j}(k)  \nn
&& - k_1  
\left(
\psi_{+,\uparrow}^\dagger(k)
\psi_{-,\uparrow}^\dagger(-k)
+\psi_{+,\downarrow}^\dagger(k)
\psi_{-,\downarrow}^\dagger(-k)
+ h.c.
\right)
\Bigg\}.
\eqa
The last term represents a pairing 
which can be written as
\bqa
 i \Bigl( \psi_{+,\uparrow}^\dagger(k), \psi_{+,\downarrow}^\dagger(k) \Bigr)
\Bigl( \vec d(k) \cdot \boldsymbol \sigma  \Bigr) 
\sigma_y
\left(
\begin{array}{c}
\psi_{-,\uparrow}^\dagger(-k) \\
\psi_{-,\downarrow}^\dagger(-k) 
\end{array}
\right) + h.c.
\eqa
with $\vec d(k) = i  k_1 \hat y$.
It is noted that $\vec d(k)$ transforms as a vector 
under $SU(2)$ spin rotations.
Therefore this describes a $p$-wave spin triplet superconducting state.
Without the pairing term, 
one has the cylindrical Fermi surface 
with co-dimension one at $\pm k_2 + \sqrt{2} k_3^2 = 0$.
The pairing gaps out the Fermi surface 
except for the line node at $k_1=0$.
This is illustrated in Fig. \ref{fig:FS_3d}.
The triplet pairing breaks the $U(1)$ symmetry associated with the fermion number conservation to $Z_2$
and the $SU(2)$ spin rotational symmetry to $U(1)$.
The theories in general dimensions 
continuously interpolate the Fermi surface in two dimensions
to the triplet superconducting state in three dimensions.
In terms of symmetry, the theory in $d=2$ is a 
special point with an enhanced symmetry.

\section{Renormalization Group}

\subsection{Minimal Action}

To start with a renormalization group analysis,
we first focus on the quadratic action in Eq.~(\ref{act4}). 
The leading terms in the quadratic action 
are invariant under the scale transformation,
\begin{eqnarray}
\vec K & = & \frac{\vec K'}{b}, \label{s1} \\
k_{d-1} & = & \frac{k_{d-1}'}{b}, \label{s2} \\
k_d & = & \frac{k_d'}{\sqrt{b}}, \label{s3} \\
\Psi (k) & = & b^{\frac{d}{2}+\frac{3}{4}} \Psi' (k'), \label{s4} \\
\phi (k) & = & b^{\frac{d}{2}+\frac{3}{4}} \phi'(k'). \label{s5} 
\end{eqnarray}
Under the scaling with $b>1$, 
the fermion-boson coupling scales as 
\begin{equation}
e' = b^{\frac12 (\frac52 -d)} e.
\end{equation}
The coupling $e$ is irrelevant for $d>d_{{\rm cr}}=5/2$, 
and is relevant for $d<5/2$.
This allows  one to access an interacting non-Fermi liquid state 
perturbatively in $d = 5/2 -\epsilon$ 
using $\epsilon$ as a small parameter.

We note that 
$|\vec K|^2 \phi^* (k)\phi(k)$ 
and
$k_{d-1}^2 \phi^* (k)\phi(k)$ 
are irrelevant in the low energy limit. 
Only $k_d^2 \phi^* (k)\phi(k)$ 
is kept in the quadratic action of the boson.
The frequency dependent self energy of the boson will
be dynamically generated by particle-hole fluctuations.
Therefore, the minimal local action is given by
\begin{eqnarray}\label{act5}
S & = &  \sum_{j} \int \frac{d^{d+1}k}{(2\pi)^{d+1}} \bar \Psi_j(k)
\Bigl[ 
i \vec \Gamma \cdot \vec K  
+ i \gamma_{d-1} \delta_k \Bigr] \Psi_{j}(k) \nonumber\\
&+&
\frac{1}{2} \int  \frac{d^{d+1}k}{(2\pi)^{d+1}} ~~
   k_d^2  \phi(-k) \phi(k) \nonumber \\
 &+&     \frac{ie \mu^{\epsilon/2}}{\sqrt{N}}\sqrt{d-1} \sum_{j}  
\int \frac{d^{d+1}k d^{d+1}q}{(2\pi)^{2d+2}}  
\phi(q) \bar \Psi_{j}(k+q) \gamma_{d-1} \Psi_{j}(k),
\end{eqnarray}
where a mass scale $\mu$ is introduced
to make the coupling constant dimensionless.
Short-ranged four fermion interactions 
and the $\phi^4$ term for the boson are not included 
in the minimal action because they are irrelevant near $d=5/2$.
There is no further BCS instability 
that gaps out the line nodes near $d=5/2$
because the density of state vanishes at Fermi energy.

For $d > 5/2$, the interaction is irrelevant 
and the low energy physics is governed by the scaling
in Eqs. (\ref{s1})-(\ref{s5}) 
with the dynamical critical exponent $z=1$.
For $d < 5/2$, the scaling will be modified such that
the interaction plays the dominant role.
In order to see this, one can choose an alternative scaling, 
\begin{eqnarray}
\vec K & = & \frac{\vec K'}{b^z}, \label{s21} \\
k_{d-1} & = & \frac{k_{d-1}'}{b}, \label{s22} \\
k_d & = & \frac{k_d'}{\sqrt{b}}, \label{s23} \\
\Psi (k) & = & b^{\frac{(d-1)z}{2}+\frac{5}{4}} \Psi' (k'), \label{s24} \\
\phi (k) & = & b^{\frac{(d-1)z}{2}+\frac{5}{4}} \phi'(k'). \label{s25} 
\end{eqnarray}
The condition that the interaction is kept marginal 
at the expense of making 
the first quadratic term in Eq. (\ref{act5}) irrelevant
uniquely fixes the dynamical critical exponent to be
$z = \frac{3}{3- 2\epsilon}$ at the tree level.
As will be shown later, 
this is indeed what we obtain from a full-fledged computation
modulo a correction coming from
the anomalous dimension of the boson field.

\subsection{Symmetry}
\label{sec:symm}

In this section, we discuss about the symmetry
of the action in Eq. (\ref{act5}). 
The $(2+1)$-dimensional theory in Eq. (\ref{act1})
has two classical $U(N)$ symmetries given by
\bqa
U_A(N) : 
\Psi_i & \rightarrow & [ e^{i \theta_\alpha T^\alpha } ]_{ij} \Psi_j, \label{u11} \\
U_B(N) : 
\Psi_i & \rightarrow & [ e^{i \sigma_z \varphi_\alpha T^\alpha } ]_{ij} \Psi_j, \label{u12} 
\eqa
where $T^\alpha$ with $\alpha = 1, 2, .., N^2$ represent
$N \times N$ Hermitian matrices. 
$U_A(N)$ and $U_B(N)$ respectively 
correspond to the global and axial symmetry groups
of the underlying $(1+1)$-dimensional theory
when $k_2$ is interpreted as an internal flavor.
The generators of the two $U(N)$ groups are given by
\bqa
j_{A0}^\alpha & = & 
\psi_{+,i}^\dagger T_{ij}^\alpha \psi_{+,j}
- \psi_{-,i}^\dagger (T^{\alpha})^{T}_{ij} \psi_{-,j}, \label{jA0} \\
j_{B0}^\alpha & = & 
\psi_{+,i}^\dagger T^\alpha_{ij} \psi_{+,j}
+ \psi_{-,i}^\dagger (T^{\alpha})^{T}_{ij} \psi_{-,j}, \label{jB0}
\eqa
where $(T^{\alpha})^{T}$ denotes the transpose of $T^\alpha$.
In general dimensions,  
only the $U_A(N)$ symmetry is kept.
The axial $U_B(N)$ symmetry is absent in $d>2$ 
because fermions with opposite chiralities are mixed.
The inability to keep both symmetries is related to
the chiral anomaly in $(1+1)$ dimensions.

The theory in general dimensions retain 
the Ward identity and the sliding symmetry
of the $(2+1)$-dimensional theory\cite{metlsach1}.
In the Ising-nematic case, 
the boson couples to the $(d-1)$-th component of the 
$U_A(1)$ current.
This implies the Ward identity
\bqa
\Gamma(k,0) & = &  \frac{ \partial G^{-1}(k) }{\partial k_{d-1} },
\label{WI}
\eqa
where $\Gamma(k,q)$ is the fermion-boson vertex function,
and $G(k)$ is the fermion propagator.
The theory also has the sliding symmetry
along the Fermi surface given by
\bqa
\Psi( \vec K, k_{d-1}, k_d ) & \rightarrow & 
\Psi( \vec K, k_{d-1} - \sqrt{d-1}( 2 \theta k_d + \theta^2), k_d + \theta ), \nn
\phi( \vec Q, q_{d-1}, q_d ) & \rightarrow & 
\phi( \vec Q, q_{d-1} - 2 \sqrt{d-1} \theta q_d, q_d ).
\eqa
As a result, the fermion propagator
depends on $k_{d-1}$ and $k_d$ only through $\delta_k$,
and the boson propagator is independent of $q_{d-1}$,
\bqa
G( \vec K, k_{d-1}, k_d ) & = & G( \vec K, \delta_k ), \nn
D( \vec Q, q_{d-1}, q_d ) & = & D( \vec Q, q_d ).
\eqa
Finally, the action respects the 
$(d-1)$-dimensional rotational symmetry 
in the space of $\vec K$
and the time-reversal symmetry.

\subsection{Renormalization Group Equation}

\begin{figure}
\begin{center}
        \includegraphics[height=4cm,width=8cm]{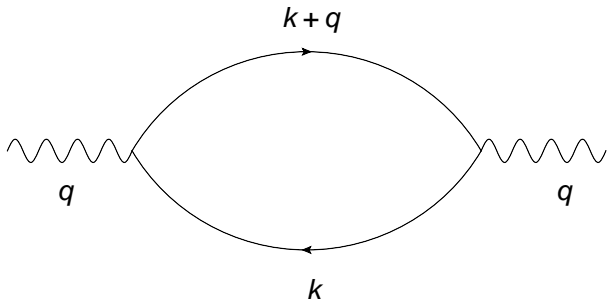} 
\end{center}
\caption{The one-loop boson self-energy.}
\label{fig:RPA}
\end{figure}

We use the field theoretical renormalization group approach 
to study the scaling behaviour of the theory 
in $d<5/2$, using $\epsilon=\frac52 -d$ as a perturbative parameter. 
At each order in the loop expansion,  
we add counter terms to cancel 
divergent terms in $1/\epsilon$
using the minimal subtraction scheme.
The bare propagator for fermions is given by
\beq\label{propf}
G_0 (k) =\frac{1}{i} \, \frac{\vec \Gamma \cdot \vec K +
\gamma_{d-1} \delta_k} 
{\vec K^2  + \delta_k^2}.
\eeq
Since the bare kinetic term of boson depends only on $k_d$,
one has to include the lowest order quantum correction 
to ensure IR and UV finiteness.
Therefore, we use the dressed propagator for boson
which includes the one-loop self-energy 
as is shown in Fig.~\ref{fig:RPA},
\beq\label{babos}
D_1(k) = \frac{1}{k_d^2 - \Pi_1 (k)} = 
\frac{1}{ k_d^2  + \beta_d e^2 \mu^{\epsilon} \frac{|\vec K|^{d-1}}{|k_d|} },
\eeq
where
\beq\label{betad}
\beta_d = \frac{\sqrt{d-1} \Gamma^2 (d/2)}
{2^d \pi^{(d-1)/2}\, |\cos(\pi d/2)|\, \Gamma(\frac{d-1}{2}) \Gamma (d)}.
\eeq
We use the sign convention where the self energy subtract the bare action
in the dressed propagator as $D(k)=\frac{1}{D_1^{-1} (k) -\Pi (k)}$,
$G(k)=\frac{1}{G_0^{-1} (k) -\Sigma (k)}$,
where $\Pi(k)$ and $\Sigma(k)$ are the self energies of
boson and fermion respectively.
The one-loop boson self-energy $\Pi_1 (k)$ is finite 
for $2\le d < 3$.
At $d=5/2$, it has the same scaling dimension as $k_d^2$ as expected. 
For computation of $\Pi_1(k)$, see Appendix~\ref{app:oneloopbos}. 

We note that the inclusion of the one-loop boson self energy
in the zero-th order quantum effective action 
is nothing but a rearrangement in the perturbative expansion
of a local theory.
This is because the non-local self energy 
is dynamically generated
from the local action.
The fact that  the one-loop boson self energy has to be included 
from the beginning has some consequences.
First, the `loop-expansion' we are going to use is defined
modulo the inclusion of the one-loop self energy of boson.
For examples, the diagrams in Fig. \ref{fig:fermvert} are regarded
as one-loop diagrams although the boson propagators in the diagrams
already include the RPA sum of boson self energy.
Second, the dynamics of boson has a intrinsic crossover scale
at $k_d \sim e^{2/3} |\vec K|^{(d-1)/3}$ which goes to zero
in the weak coupling limit.
Because of this, the actual parameter that controls the loop expansion
is not $e$ as will be discussed in Sec. III D in more detail.

The counter terms take the same form 
as the original local action,
\begin{eqnarray}\label{act6}
S_{CT} & = &  \sum_{j} \int \frac{d^{d+1}k}{(2\pi)^{d+1}} \bar \Psi_j(k)
\Bigl[ 
i A_1 \vec \Gamma \cdot \vec K 
+ i A_2 \gamma_{d-1} \delta_k 
 \Bigr] \Psi_{j}(k) 
\nonumber\\
&+&
\frac{A_3}{2} \int  \frac{d^{d+1}k}{(2\pi)^{d+1}}~~
k_d^2 \phi(-k) \phi(k) \nonumber \\
 &+&    A_4 \frac{ie \mu^{\epsilon/2} }{\sqrt{N}}\sqrt{d-1} \sum_{j}  
\int \frac{d^{d+1}k d^{d+1}q}{(2\pi)^{2d+2}}  
\phi(q) \bar \Psi_{j}(k+q) 
\gamma_{d-1} \Psi_{j}(k),
\end{eqnarray}
where 
\bq
A_n = \sum_{k=1}^\infty \frac{Z_{n,k}(e)}{\epsilon^k}.
\eq
In the mass independent minimal subtraction scheme,
the coefficients $Z_{n,k}(e)$ depend only on the coupling.
$A_n$ and $Z_{n,k}$ can be further expanded in the number of loops.
We use $A_n^{(L)}$ and $Z_{n,k}^{(L)}$ to denote $L$-loop contributions
modulo the one-loop self energy of boson which 
is already included in Eq. (\ref{babos}).
Note that the $(d-1)$-dimensional rotational invariance
in the space perpendicular to the Fermi surface
guarantees that  $\Gamma_i K_i$ 
are renormalized in the same way
for $0\leq i \leq (d-2)$.
Similarly, the sliding symmetry along the Fermi surface
guarantees that the form of $\delta_k$ is preserved.
However, $A_1$ and $A_2$ are in general different
due to a lack of the full rotational symmetry in the $(d+1)$-dimensional spacetime.
This will leads to a non-trivial dynamical critical exponent 
as will be shown later.
The Ward identity in Eq. (\ref{WI}) forces $A_4 = A_2$.

Adding the counter terms to the original action,
we obtain the renormalized action
which gives the finite quantum effective action,
\begin{eqnarray}\label{act7}
S_{ren} & = &  \sum_{j} \int \frac{d^{d+1}k_B}{(2\pi)^{d+1}} 
\bar \Psi_{Bj}(k_B)
\Bigl[ 
i  \vec \Gamma \cdot \vec K_B + 
i \gamma_{d-1} \delta_{k_B}  \Bigr] \Psi_{Bj}(k_B) 
\nonumber\\
&+&
\frac{1}{2} \int  \frac{d^{d+1}k_B}{(2\pi)^{d+1}} ~
k_{Bd}^2 \phi_B(-k_B) \phi_B(k_B) \nonumber \\
 &+&     \frac{ie_B }{\sqrt{N}}\sqrt{d-1} \sum_{j}  
\int \frac{d^{d+1}k_B d^{d+1}q_B}{(2\pi)^{2d+2}}  
\phi_B(q_B) \bar \Psi_{Bj}(k_B+q_B) 
\gamma_{d-1} \Psi_{Bj}(k_B),
\end{eqnarray}
where
\bqa
\vec K & = & \frac{Z_2}{Z_1} \vec K_B, \nn
k_{d-1} & = & k_{B, d-1}, \nn
k_d & = & k_{B,d}, \nn
\Psi(k) & = &  Z_\Psi^{-1/2} \Psi_B(k_B), \nn
\phi(k) & = & Z_\phi^{-1/2} \phi_B(k_B), \nn
e_B & = & Z_3^{-1/2} \left( \frac{Z_2}{Z_1} \right)^{(d-1)/2} \mu^{\epsilon/2} e
\eqa
with $Z_n  =  1 + A_n$,
$Z_\Psi =  Z_2 \left( \frac{Z_2}{Z_1} \right)^{(d-1)}$ and
$Z_\phi = Z_3 \left( \frac{Z_2}{Z_1} \right)^{(d-1)}$.
In Eq. (\ref{act7}), there is a freedom to change 
the renormalizations of the fields 
and the renormalization of momentum
without affecting the action.
Here we fix the freedom by requiring 
that  $\delta_{k_B} = \delta_k$.
This amounts to measuring scaling dimensions 
of all other quantities 
relative to that of $\delta_k$.

The finite renormalized Green's function 
is defined by
\bqa
&& \Bigl<
\bar
\Psi( k_1 )  .. \bar \Psi(k_m) ~
\Psi( k_{m+1} )  .. \Psi(k_{2m}) ~
\phi( k_{2m+1} ) .. \phi(k_{2m+n})
\Bigr> \nn
&& = G^{(m,m,n)}( \{ k_i \}; e, \mu ) ~~
\delta^{d+1} \left( \sum_{i=1}^{m} k_i  - \sum_{j=m+1}^{2m+n} k_j \right),
\eqa
where the flavor and spacetime indices of fermions 
are suppressed.
It is related to the bare Green's function defined by
\bqa
&& \Bigl<
\bar \Psi_B( k_{B1} )  .. \bar \Psi_B(k_{Bm}) ~
\Psi_B( k_{B m+1} )  .. \Psi_B(k_{B 2m}) ~
\phi_B( k_{B 2m+1} ) .. \phi_B(k_{B 2m+n})
\Bigr> \nn
&& = G_B^{(m,m,n)}( \{ k_{B i} \}; e_B ) ~~
\delta^{d+1} \left( \sum_{i=1}^{m} k_{B i}  - \sum_{j=m+1}^{2m+n} k_{B j} \right),
\eqa
through the multiplicative renormalization,
\bqa
G^{(m,m,n)}(\{ k_i \}; e, \mu ) 
= 
Z_\Psi^{-m}
Z_\phi^{-\frac{n}{2}}
\left( \frac{Z_2}{Z_1} \right)^{d-1}
G_B^{(m,m,n)}( \{ k_{Bi} \}; e_B ). 
\eqa
Using the facts that the bare Green's function
is independent of $\mu$
and that $G^{(m,m,n)}$ has the engineering 
scaling dimension $-(2m+n)\frac{4-\epsilon}{2}+(3-\epsilon)$,
one obtains the renormalization group equation,
\bqa
&&
\Bigg\{
\sum_{i=1}^{2m+n} \left(
z(e)\, \vec K_i \cdot \nabla_{K_i}
+ k_{i, d-1} \frac{\partial}{\partial k_{i,d-1}}
+ \frac{ k_{i,d} }{2} \frac{\partial}{\partial k_{i,d}} 
\right)
 - \beta(e) \frac{\partial}{\partial e} 
 - 2m \left[ -\frac{4-\epsilon}{2} +  \eta_\psi \right] \nn
&& - n \left[ -\frac{4-\epsilon}{2} +  \eta_\phi \right]
- \left[ z(e) \left( \frac{3}{2} -\epsilon \right) + \frac{3}{2} \right]
\Bigg\}
G^{(m,m,n)}(\{ k_i \}; e, \mu ) 
=0.
\label{RGeq}
\eqa
From the equation, 
it is clear that the dimension of $\vec K$ is renormalized to $z$, 
and the total dimension of spacetime becomes 
$-\left[ z \left( \frac{3}{2} -\epsilon \right) + \frac{3}{2} \right]$
accordingly.
Here  $z$ is the dynamical critical exponent,
$\beta$ is the beta function,
and $\eta_\psi$ ($\eta_\phi$) is 
the anomalous dimensions for fermion (boson)
which are given by
\bqa
z(e) & = & 1 - \frac{ \partial \ln (Z_2/Z_1) }{\partial \ln \mu}, \nn
\beta(e) & = &  \frac{\partial e}{\partial \ln \mu}, \nn
\eta_\psi(e) & = & \frac{1}{2} \frac{ \partial \ln Z_\Psi}{\partial \ln \mu}, \nn
\eta_\phi(e) & =& \frac{1}{2} \frac{ \partial \ln Z_\phi}{\partial \ln \mu}.
\label{beta1}
\eqa
We use the convention that the beta function describes
the flow of the coupling with increasing energy scale.
These four equations can be rewritten as
\bqa
\beta ( Z_1 Z_2^{'} - Z_2 Z_1^{'} ) + Z_1 Z_2 (z-1) & = & 0, \nn
e \left[ -\frac{\epsilon}{2} z + \frac{3}{4} (z-1) \right] Z_3
- \left[ Z_3 - \frac{e}{2} Z_3^{'} \right] \beta & = & 0, \nn
Z_2 \eta_\psi - \frac{\beta}{2} Z_2^{'} + \left( \frac{3}{4} - \frac{\epsilon}{2} \right) (z-1) Z_2 & = & 0, \nn
Z_3 \eta_\phi - \frac{\beta}{2} Z_3^{'} + \left( \frac{3}{4} - \frac{\epsilon}{2} \right) (z-1) Z_3 & = & 0,
\label{beta2}
\eqa
where primes represent derivatives with respect to $e$.
One can readily see that 
the regular part of Eqs. (\ref{beta1})
in the $\epsilon \rightarrow 0$ limit 
requires the solutions of the form,
\bqa
z & = & z^{(0)}, \nn
\beta & =  & \beta^{(1)} \epsilon + \beta^{(0)}, \nn
\eta_\psi & = & \eta_{\Psi}^{(1)} \epsilon + \eta_{\Psi}^{(0)}, \nn
\eta_\phi & = & \eta_{\phi}^{(1)} \epsilon + \eta_{\phi}^{(0)}.
\eqa
Using this form, 
one can solve Eqs. (\ref{beta2}) at each order in $\epsilon$.
$z, \beta, \eta_\psi, \eta_\phi$ 
are determined from the simple poles of the counter terms as
\bqa
z & = & \frac{2}{ 2 + e ( Z_{1,1}^{'} - Z_{2,1}^{'} ) }, \label{zcal} \\
\beta & = & 
\left( 
-\frac{\epsilon }{2} z    + \frac{3}{4} ( z - 1 ) 
\right) e 
- \frac{z e^2}{4} Z_{3,1}^{'}, \label{bcal} \\
\eta_\psi & = & -\frac{( z - 1) (3-2\epsilon)}{4} - \frac{z e}{4}  Z_{2,1}^{'}, \label{psical} \\
\eta_\phi & = & -\frac{( z - 1) (3-2\epsilon)}{4} - 
\frac{z e}{4}  Z_{3,1}^{'}. \label{phical}
\eqa
In Eq. (\ref{zcal}), we see that the dynamical critical exponent 
is renormalized by quantum effects.
In other words, the first $(d-1)$ components 
of the energy-momentum vector
acquires an anomalous dimension $(z-1)$.
The anomalous dimension of spacetime
affects the scaling dimension of the coupling
and the anomalous dimensions of the fields
in Eqs. (\ref{bcal}), (\ref{psical}) and (\ref{phical}).
Once $Z_{n,m}^{'}$ are computed,
one can obtain the beta functions
and the critical exponents.

The theory has the Gaussian fixed point at which
$e=0, z=1, \eta_\psi= \eta_\phi = 0$.
As a small coupling is turned on,
the theory flows to an interacting fixed point with $e^* \neq 0$ at low energies.
The condition that the beta function vanishes at the interacting fixed point
determines the dynamical critical exponent to be
\bqa
z^* = \frac{3}{3 - 2 \epsilon - e^* Z_{3,1}^{'}}.
\eqa
It is remarkable that the dynamical critical exponent at the fixed point 
is independent of $Z_1$ and $Z_2$.
If $Z_{3,1}=0$, $z^*$ is exactly given by
$z^* = \frac{3}{3 - 2 \epsilon }$ 
which monotonically increases from $1$ to $3/2$  
as $d$ changes from $5/2$ to $2$.
At the fixed point,
the scaling dictates the form of the two-point functions as 
\bqa
D(k) & = & \frac{1}{ ( k_d^2 )^{1 - (z-1)(3/2-\epsilon) - 2 \eta_\phi} }
~~ f \left( \frac{| \vec K |^{1/z} }{k_d^2} \right),  \label{Dk0} \\
G(k) & = & \frac{1}{ | \delta_k |^{1 - (z-1)(3/2-\epsilon) - 2 \eta_\psi} }
~~ g \left( \frac{ | \vec K |^{1/z} }{\delta_k} \right), \label{Gk0}
\eqa
where $f(x)$ and $g(x)$ are universal cross-over functions.
The flavor and the spinor indices are suppressed in $g(x)$.
If the anomalous dimensions are large enough, 
the singularity in the Green’s functions can in principle turn 
into an algebraic gap\cite{UNP}. 
However, the anomalous dimensions are small near the
upper critical dimension.

\subsection{Expansion parameter}

We take the small $\epsilon$ limit with fixed $N$.
In this section, we show that the loop expansion is controlled in this limit.
Although the bare fermion-boson vertex includes $e$,
it is not the actual expansion parameter. 
This is due to the fact that the boson propagator
includes the self energy which vanishes
in the $e \rightarrow 0 $ limit.
To examine this issue more closely,
let us consider a boson propagator 
which carries an internal momentum $k$
within a diagram.
The integration over $k$ is of the form,
\bqa
\int dk ~ 
F( k, \{ q_i \} ) \frac{1}{ k_d^2  + \beta_d e^2 \mu^\epsilon
\frac{|\vec K|^{d-1}}{|k_d|} },
\eqa
where $\{ q_i \}$ is a set of other internal and external momenta,
and $F( k, \{ q_i \} )$ represents the contribution from other propagators.
When $k_d$ can be arbitrarily small in magnitude,
the integration is in general IR divergent 
in the $e \rightarrow 0$ limit.
The IR divergence is cut-off at a scale 
$k_d \sim e^{2/3} |\vec K|^{(d-1)/3}$,
and the result of the integration becomes order of $e^{-2/3}$.
Therefore, each boson propagator contributes 
an IR enhancement factor of $e^{-2/3}$
provided that the internal momentum
that runs through each boson propagator 
is allowed to vanish {\it independently}.

If there is a kinematic constraint that keeps $k_d$
from becoming arbitrarily small in magnitude,
$k_d$ integration is convergent in the $e \rightarrow 0$ limit. 
Then there is no IR enhancement factor.
However, one still has to worry about UV divergence
in the $e \rightarrow 0$ limit.
In particular, the integration over $\vec K$ can be UV divergent
without the self energy in the boson propagator.
In the presence of the self-energy,
quantum corrections to marginal operators
can have at most log divergences by power counting.
In the $e \rightarrow 0$ limit, 
they can have power-law UV divergences
because the boson propagator no longer depends on $\vec K$.
The degree of UV divergence for marginal operators
is {\it at most}  $I_b$,
where $I_b$ is the number of internal boson propagators.
This is because only the boson propagator depends on $e$,
and each boson propagator carries 
the scaling dimension $-1$.
In the presence of the boson self energy,
the power-law divergence is cut-off at a scale
$|\vec K| \sim e^{-2/(d-1)} k_d^{3/(d-1)}$.
In $d=5/2$, the UV divergence in the $e = 0$ limit 
can introduce an enhancement factor of $e^{-4/3 I_b}$.
However, we emphasize that this is an upper bound for the 
enhancement factor.
Typical diagrams have weaker UV divergence
in the $e \rightarrow 0$ limit due to kinematic constraints,
which results in a smaller enhancement factor.

In the presence of the IR and UV enhancement factors,
a $L$-loop diagram goes as
\bqa
e^{2L - Y I_b} = e^{ ( 2 - Y ) L - ( E_f/2 - 1) Y  }.
\label{e}
\eqa
Here $2/3 \leq Y \leq 4/3$ is the average enhancement factor per boson propagator,
which is specific to each diagram.
The identity $I_b = L - 1 + E_f/2$ is used, 
where $E_f$ is the number of external fermion lines.
From explicit calculations, 
we will see that all diagrams up to two-loop level
have $Y=2/3$. 
At the three-loop order, we will see an example where $Y=1$.
At present, we don't have any example with $Y>1$.
Up To the three-loop diagrams that we have checked,
all $L$-loop diagrams are suppressed by $e^{4/3 L}$,
compared to the bare action and the one-loop self energy of boson.
This suggests that 
the actual average enhancement factor 
may be strictly smaller than $4/3$.
Although we don't know the precise expansion parameter,
all $L$-loop diagrams are suppressed at least by
the factor of $e^{ 2/3 L}$, 
and the loop expansion is controlled.

\subsection{Computation of counter terms}

In this section, we summarize the results of the counter terms
computed up to two loops.
Some three-loop diagrams are also computed.

\subsubsection{One-loop level}

\begin{figure}
\begin{center}
\includegraphics[height=4cm,width=8cm]{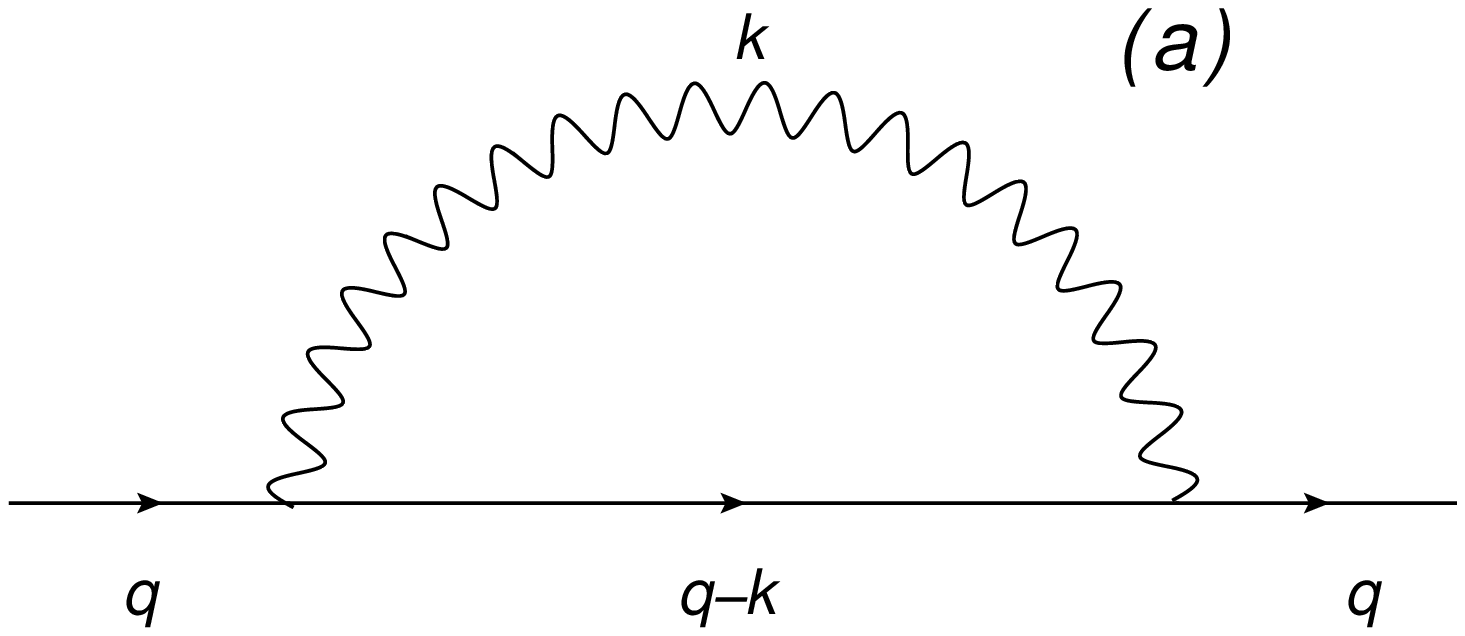}
\end{center}
\vspace{0.7cm}
\begin{center} 
\includegraphics[height=4cm,width=8cm]{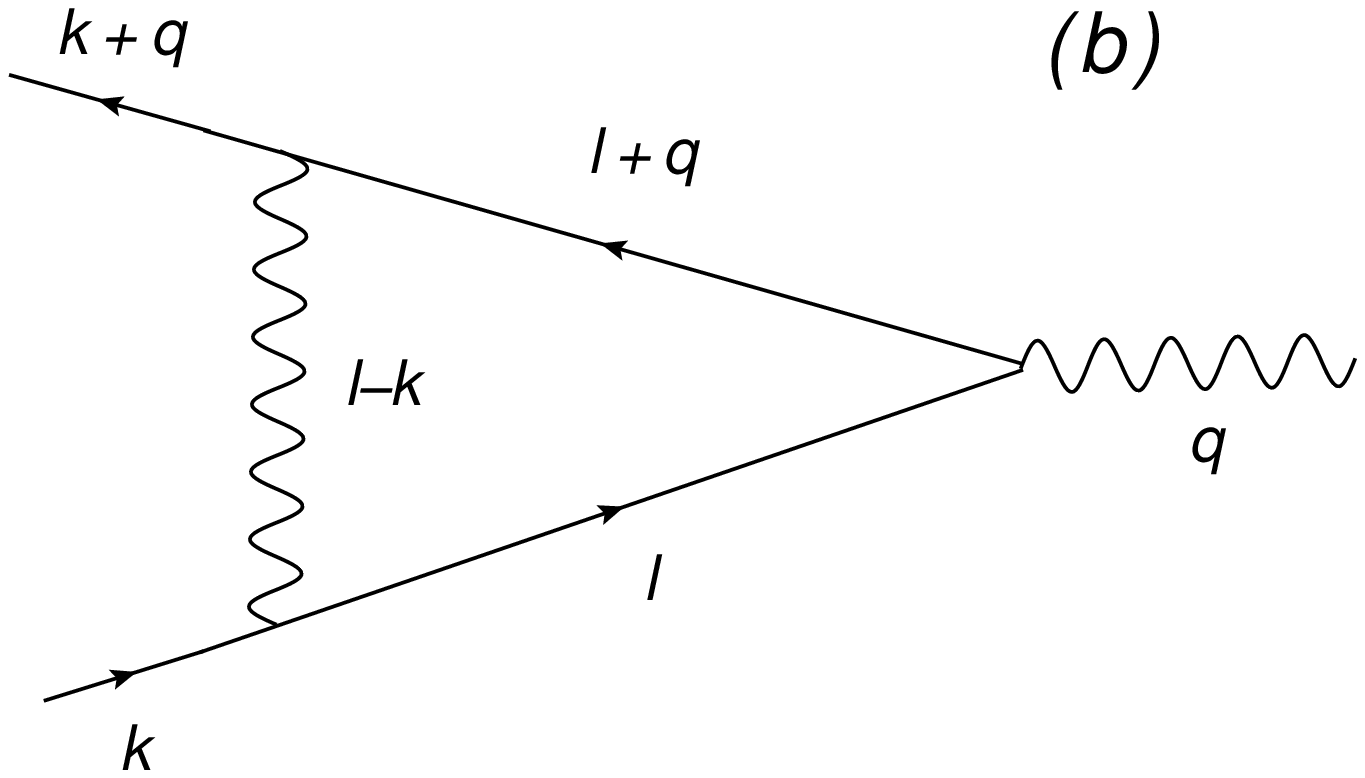} 
\end{center}
\caption{(a) The one-loop fermion self-energy.  
(b) The one-loop vertex correction. }
\label{fig:fermvert}
\end{figure}

The one-loop self energy of boson has been 
already taken into account in the dressed propagator, $D_1(k)$.
The one-loop fermion self energy shown in 
Fig.~\ref{fig:fermvert} (a) is given by
\beq\label{fermiloop}
\Sigma_1 (q) = \frac{(ie)^2 \mu^{\epsilon}}{N} (d-1) 
\int \frac{dk}{(2\pi)^{d+1}} 
\gamma_{d-1} G_0 (q-k)\gamma_{d-1} D_1 (k),
\eeq  
where 
\beq\label{propbos}
D_1 (k) =\frac{1}{k_d^2 +\beta_{d} e^2 \mu^{\epsilon}
\displaystyle\frac{|\vec K|^{d-1}}{|k_d|} }.
\eeq
As is computed in Appendix~\ref{app:oneloopferm},
the resulting self energy is given by
\beq
\Sigma_1(q) =
\left(  - \frac{e^{4/3}}{N} \frac{u_1}{\epsilon} + \mbox{finite terms}  \right)
(i \vec \Gamma \cdot \vec Q ) 
\eeq
with 
\bqa
u_1 = \frac{1}{ 2^{5/2} \pi^{3/4} \sqrt{3} \beta_{5/2}^{1/3} \Gamma(3/4) } 
& \approx  & 0.08758634.
\eqa
It is noted that 
not only the UV divergent part but also the finite part in 
$\Sigma_1(k)$ is proportional to $\vec \Gamma \cdot \vec Q$.
This fact simplifies the calculation at higher loops as will be discussed in the next section.
To cancel the UV divergence, 
we only need the counter term of the form,
\beq\label{actct}
S_{CT}^{(1loop)}  =  \sum_{j} \int \frac{dk}{(2\pi)^{d+1}} \bar \Psi_j(k)\,
i A_1^{(1)} 
(\vec \Gamma \cdot \vec K )
\, \Psi_{j}(k)
\eeq
with
\beq
A_1^{(1)}  = - \frac{e^{4/3}}{N} \frac{u_1}{\epsilon}.
\eeq

The absence of $\delta_k$ dependence in the fermion self energy 
combined with the Ward identity in Eq. (\ref{WI}) 
implies that there is no vertex correction
at the one-loop.
This is explicitly checked in Appendix~\ref{app:oneloopvert}.

\subsubsection{Two-loop level}

\begin{figure}
\begin{center}
\includegraphics[width=15cm]{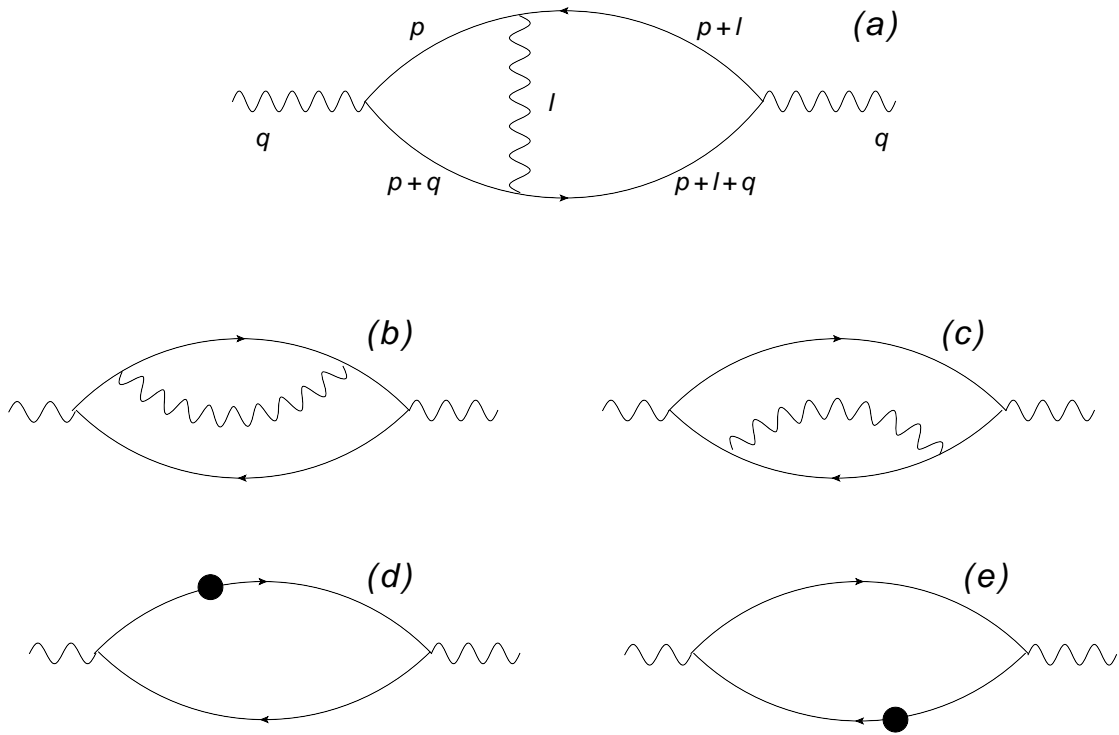}
\end{center}
\caption{The diagrams for two-loop boson self energy.}
\label{fig:bos2}
\end{figure}
\begin{figure}
\begin{center}
\includegraphics[width=15cm]{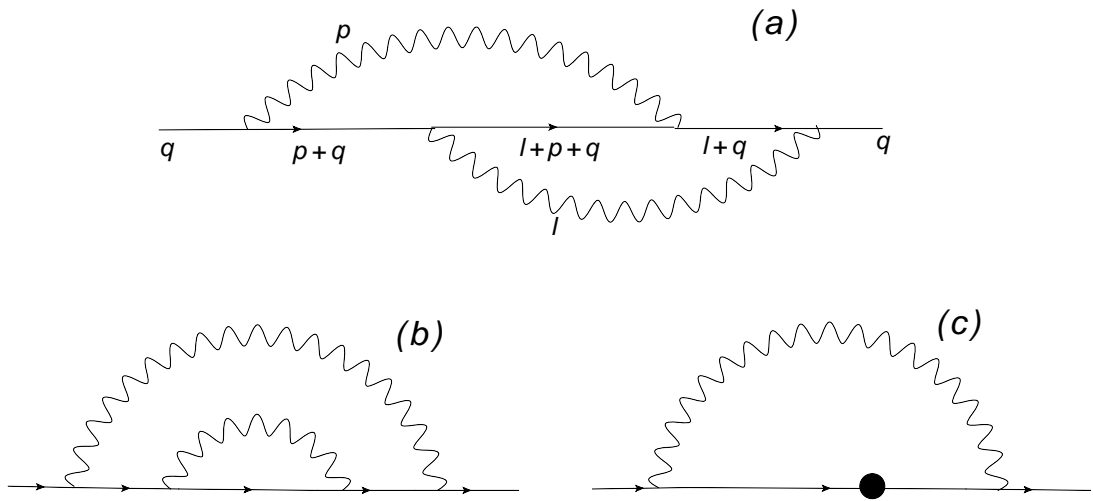}
\end{center}
\caption{The diagrams for two-loop fermion self energy.}
\label{fig:ferm2}
\end{figure}
\begin{figure}
\begin{center}
\includegraphics[width=15cm]{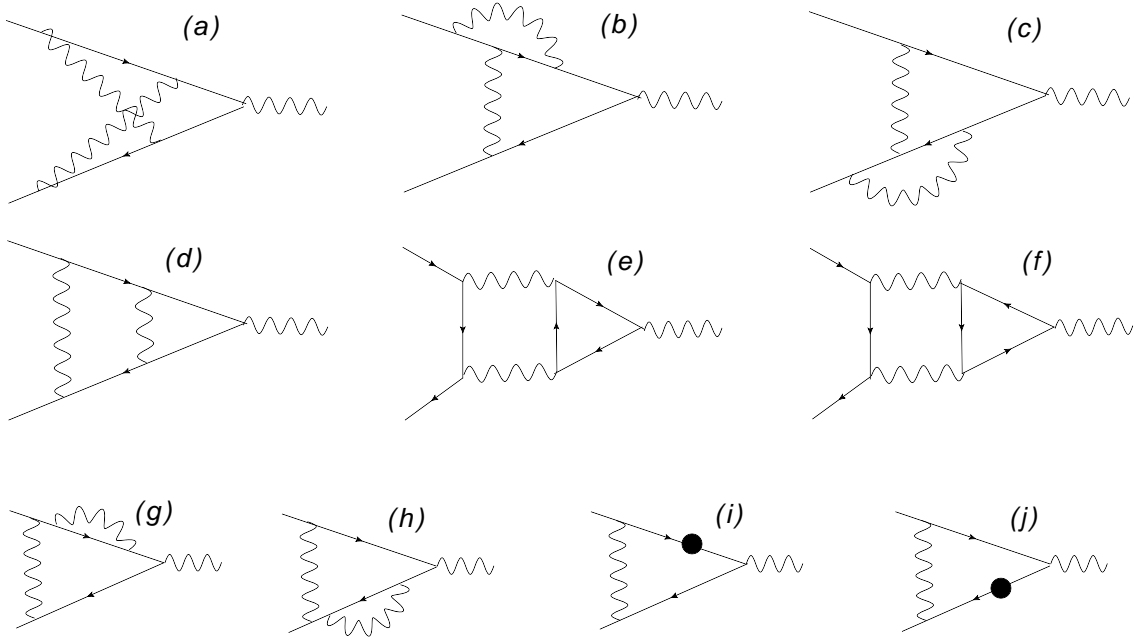}
\end{center}
\caption{The diagrams for two-loop vertex corrections.}
\label{fig:vert2}
\end{figure}

The two-loop diagrams are listed in 
Figs.~\ref{fig:bos2} ,\ref{fig:ferm2} and~ \ref{fig:vert2}.  
The black circles in  
Figs. \ref{fig:bos2} (d)-(e), \ref{fig:ferm2}(c) and 
\ref{fig:vert2}(i)-(j) denote the one-loop
counter term for the fermion self energy,
\beq\label{2legv}
i A_1^{(1)} \bar \Psi (\vec \Gamma \cdot \vec K ) \Psi.
\eeq        
To examine which diagrams can give non-zero contributions, 
we first note that the fermion self-energy of the form
\beq\label{eliashferm}
\Sigma (k) =-i \xi (K) [\vec \Gamma \cdot \vec K]
\eeq
with $K=|\vec K|$ 
solves the Eliashberg equations for the bosonic and 
fermionic self-energies.
If one uses
the dressed fermionic propagators
\beq\label{fullferp}
G(k) = [G_0^{-1} (k) -\Sigma (k)]^{-1}
\eeq
in lieu of the bare one, 
one obtains the same self energies, $\Sigma_1 (q)$ and $\Pi_1 (q)$
which are obtained by using $G_0$.
This can be understood from the fact 
that the dependence on $\xi (K)$ drops out
in Eqs.~(\ref{fermiloop}) and (\ref{bosloop})
once $k_{d-1}$ and $k_d$ are integrated out. 
We also note that
the full one-loop fermion self-energy in Eq.~(\ref{sigma3}) 
has the form of Eq.~(\ref{eliashferm}). 
As a result, the diagrams 
in Figs. \ref{fig:bos2}(b), (c) and \ref{fig:ferm2}(b)  
vanish  
because they can be obtained by expanding the dressed propagators 
in powers of $\Sigma_1 (q)$ in the corresponding expressions for 
the one-loop diagrams.
Since the one-loop counter term is also of the same form,
the diagrams in Figs. \ref{fig:bos2}(d)-(e) and 
\ref{fig:ferm2}(c) vanish as well.
This feature can be checked by explicit computation. 
We thus conclude that the only two-loop diagrams 
that need to be computed 
for the self-energies are those in 
Figs.~\ref{fig:bos2}(a) and \ref{fig:ferm2}(a). 
The vertex correction can be obtained from the Ward identity. 

The diagram in Fig. \ref{fig:bos2}(a) is computed in Appendix~\ref{app:twoloopbos}.
Although it is UV finite, 
it renormalizes $\beta_d$ in the boson propagator
by a finite amount, $\beta_d^{6a}  \sim O(e^{4/3}/N)$.
Once this correction is fed back to the one-loop fermion self energy in Eq. (\ref{fermiloop}),
we obtain a correction to the UV-divergent fermion self energy,
\bqa
\Sigma^{6a}_2(k) &=& -\frac{\beta_d^{6a} }{3 \beta_d} \Sigma_1(k) \nn
&=& \left(  - \frac{e^{8/3}}{N^2} \frac{u_2^{\rq{}}}{\epsilon} + \mbox{finite terms}  \right)
(i \vec \Gamma \cdot \vec K ), 
\eqa
 where 
 \bqa
u_2^{\rq{}}  \approx 0.0016449.
\eqa

The two-loop fermion self-energy 
in Fig. \ref{fig:ferm2}(a) 
is given by 
\beq\label{twoloopf}
\Sigma^{7a}_2 (q) = \frac{(ie)^4 \mu^{2\epsilon}}{N^2} (d-1)^2 
\int \frac{dp dl}{(2\pi)^{2d+2}} D_1 (p) D_1 (l) 
\gamma_{d-1} G_0 (p+q)\gamma_{d-1} G_0 (p+l+q)\gamma_{d-1} G_0 (l+q)\gamma_{d-1}.
\eeq
The computation described in Appendix~\ref{app:twoloopferm} 
results in
\beq\label{twoloopf2}
\Sigma_2^{7a} (q) = 
-  \frac{ e^{8/3} }{N^2} u_2
 ( i \vec \Gamma \cdot \vec K )
-  \frac{ e^{8/3} }{N^2} v_2  ( i \gamma_{d-1} \delta_k )
+ \mbox{ finite terms},
\eeq
where
\bqa
u_2 &  \approx & - 0.0194218 \nn
v_2 &  \approx & 0.000867775.
\eqa


From the Ward identity, one has to include the vertex correction
at the two-loop level.
The counter terms that are necessary 
to cancel the UV divergence at the two-loop level is given by
\bqa\label{actct2}
S_{CT}^{(2loop)}  &=&  \sum_{j} \int \frac{dk}{(2\pi)^{d+1}} \bar \Psi_j(k)\,
[i A_1^{(2)}( \vec \Gamma \cdot \vec K )+ i A_2^{(2)}  
\gamma_{d-1} \delta_k ]\, \Psi_{j}(k) +\nonumber \\
 &+&    A_2^{(2)} \frac{ie \mu^{\epsilon/2} }{\sqrt{N}}\sqrt{d-1} \sum_{j}  
\int \frac{dk dq}{(2\pi)^{2d+2}}  
\phi(q) \bar \Psi_{j}(k+q) \gamma_{d-1} \Psi_{j}(k),
\eqa
where
\bqa\label{Z1and2}
A_{1}^{(2)} & = &  -  \frac{ e^{8/3} }{N^2} ( u_2 + u_2^{\rq{}} ), \nn
A_{2}^{(2)} & = &  -  \frac{ e^{8/3} }{N^2} v_2.
\eqa

\subsubsection{Three-loop Aslamazov-Larkin-type 
contribution to boson self-energy}

\begin{figure}
\begin{center}
\includegraphics[width=11cm]{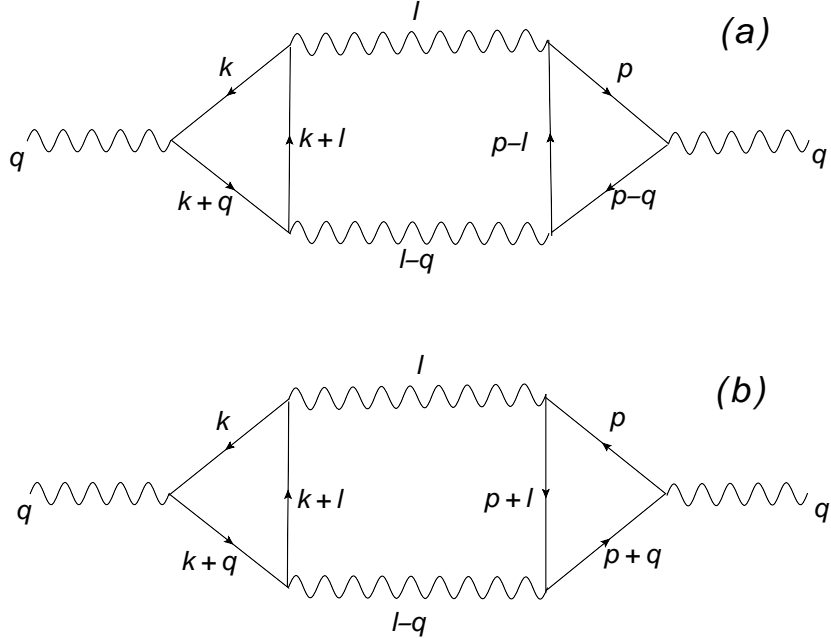}
\end{center}
\caption{Aslamazov-Larkin-type contributions to boson self-energy. 
Diagrams (a) and (b) correspond to the particle-particle and 
particle-hole channels respectively.}
\label{fig:ALbos}
\end{figure}

The number of diagrams increases dramatically at higher loops. 
This makes it hard to go beyond the two-loop level systematically.
It is of interest, however, to consider some three-loop diagrams that 
can potentially contribute to  anomalous dimension of boson through
a non-trivial correction to $Z_3$, given that $Z_3=1$ up to the two-loop order.
For this, we consider the  Aslamazov-Larkin-type diagrams shown in Fig.~\ref{fig:ALbos}
which is the lowest known diagrams that renormalize
the boson kinetic term. 
Metlitski and Sachdev evaluated these diagrams in 
Ref.~\cite{metlsach1}, 
and showed that they introduce a finite 
renormalization to the kinetic energy of bosons,
which violates the genus expansion in the two-patch theory. 
A finite quantum correction to the kinetic energy is also found by Mross et. al 
in Ref. \cite{mross}.

To extract the term that renormalizes the $q_d^2$ term in the boson action, 
we compute the diagrams at $\vec Q =0$.
Details of computation are presented in Appendix~\ref{app:ALbos}.
The final result can be written as
\bqa\label{ALres}
\Pi_{AL} (q_d) & = & q_d^2 \cdot \frac{e^6}{N} 
\left( \frac{\mu}{q_d^2} \right)^{3\epsilon} 
(d-1)^{3d/2} \int \frac{d\vec P d\vec K d\vec L}{(2\pi)^{3(d-1)}}
\frac{{\cal J}_d (L)}
{\{(P+ |\vec P +\vec L| +K +|\vec K +\vec L|)^2 +1\} } \nn
&\cdot&
\frac{\left( [\vec K \cdot (\vec K +\vec L)] -K\, |\vec K +\vec L| \right)\,
\left( [\vec P \cdot (\vec P +\vec L)] -P\, |\vec P +\vec L|
\right)}
{2 P K\,|\vec P +\vec L|\,|\vec K +\vec L| 
\left[ P+ |\vec P +\vec L| +K +|\vec K +\vec L| \right]},
\eqa
where
\bqa
{\cal J}_d (L)  &=& \int_0^1 \frac{dl_d}{2\pi} 
\frac{2^{3d-6} \, [l_d (1-l_d)]^{2(d-1)}}{ l_d^{4-d} +\beta_d e^2 
\left(\mu/q_d^2\right)^{\epsilon}\, [2\sqrt{d-1} (1-l_d) L]^{d-1} }\nn
&\cdot& \frac{1}{(1-l_d)^{4-d} +\beta_d e^2 
\left( \mu/q_d^2\right)^{\epsilon}\, [2\sqrt{d-1} l_d L]^{d-1}}.
\eqa
Here $\vec P, \vec K, \vec L$ have been rescaled to be dimensionless
in the unit of $q_d^2$.
We see that 
$\Pi_{AL} (q_d) \propto q_d^2$  
in the  $\epsilon\rightarrow 0$  limit.
One can also check that the coefficient of the $k_d^2$ term
is finite at $d=5/2$.
To see this, 
we introduce a $9/2$-dimensional vector $\vec X =(\vec L, \vec P, \vec K)$.
Since ${\cal J}_{5/2}(L)$ decays as $1/L^3$ 
in the $\vec L \rightarrow \infty$  limit, 
Eq.~(\ref{ALres}) behaves as $\int dX /X^{5/2}$ at large momenta,
which is UV convergent. 
Infrared convergence is explicit as well. 
To estimate the dependence on $e$, 
we  note that ${\cal J}_{5/2} (L)$ has a non-trivial dependence on $e$, 
and behaves differently depending on whether $L$ is large or small compared to 
$L_* =1/e^{4/3} \gg 1$ (in the unit of $q_d^2$) : 
\bqa\label{jint1} 
{\cal J}_{5/2} (L) \approx \left\{ \begin{array}{cl}
C_1 ,& \quad L \ll L_* \\
\displaystyle\frac{C_2}{e^4 L^3}, & \quad L \gg L_* , \end{array} \right.
\eqa
where $C_1$ and $C_2$ are constants
which are independent of $e$. 
It is not an easy to perform the integration over $\vec P$, $\vec K$ explicitly. 
However, based on the scaling arguments, 
one can write that
\beq\label{alint1}
\Pi_{AL} (q_d) =\frac{e^6}{N} \cdot q_d^2 \, \int dL ~ L^{1/2} {\cal F} (L) 
{\cal J}_{5/2} (L),
\eeq
where ${\cal F} (L)$ can be considered approximately constant when $L >> 1$. 
Breaking the integration into the regions $0<L<L_*$ and $L_*<L<\infty$, 
and taking into account the asymptotics given by Eq.~(\ref{jint1}), 
we obtain 
\beq
\Pi_{AL} (q_d) = C_3 \frac{e^4}{N} \cdot q_d^2.
\eeq
$C_3$ is a numerical constant independent of $e$.
Thus the Aslamazov-Larkin diagrams contribute
 a finite renormalization to the boson kinetic term.
Therefore, we still have $Z_3=1$.
It is an outstanding question 
whether $Z_3$ ever receives a non-trivial quantum correction, 
and, if so, at which order the first quantum correction appears.

\subsection{Critical exponents}

Collecting all the results, 
the counter terms up to the two-loop level are given by
\bqa
Z_{1,1} & = &  - \frac{ e^{4/3} }{N} u_1 -  \frac{ e^{8/3} }{N^2} ( u_2 + u_2^{'} ), \nn
Z_{2,1} & = &  -  \frac{ e^{8/3} }{N^2} v_2, \nn
Z_{3,1} & = & 0, \nn
\eqa
where
\bqa\label{uvnum}
u_1 & \approx  & 0.08758634, \nn
u_2 & \approx &  - 0. 0194218, \nn
u_2^{'} & \approx &   0.0016449, \nn
v_2 & \approx  & 0.000867775.
\eqa
The beta function becomes
\bqa
\beta & = & -\frac{\epsilon}{2} e
+ 0.02920 \left( \frac{3}{2} - \epsilon \right) \frac{e^{7/3}}{N}
- 0.01073 \left( \frac{3}{2} - \epsilon \right)  \frac{e^{11/3}}{N^2} 
\eqa
which has a stable interacting fixed point at
\bqa
\frac{ e^{* 4/3} }{N} & = & 11.417 \epsilon + 55.498 \epsilon^2.
\eqa
Therefore we conclude that the theory flows to a 
stable non-Fermi liquid state in the low energy limit.
To the two-loop order, the dynamical critical exponent and 
the anomalous dimensions at the critical point are given by
\bqa
z & = & \frac{3}{3 - 2 \epsilon},  \label{zcal2} \\
\eta_\psi & = & -\frac{\epsilon}{2} + 0.07541 \epsilon^2, \label{psical2} \\
\eta_\phi & = & -\frac{\epsilon}{2}, \label{phical2}
\eqa
and the propagators are given by
\bqa
D(k) & = & \frac{1}{  k_d^2 }
~~ f \left( \frac{| \vec K |^{1/z} }{k_d^2} \right), \label{Dk} \\
G(k) & = & \frac{1}{ | \delta_k |^{1-0.1508 \epsilon^2 } }
~~ g \left( \frac{ | \vec K |^{1/z} }{\delta_k} \right). \label{Gk}
\eqa
It is noted that the contribution of the dynamical critical exponent
to the anomalous dimensions of the fields,
that is the first term in Eqs. (\ref{psical}) and (\ref{phical}),
drops out in the two-point functions because of the cancellation
with the dynamical critical exponent in the delta function 
which enforces the energy-momentum conservation
in the Green's functions.
However, the contribution of the dynamical critical exponent
shows up in higher point functions.

The upper bound on the enhancement factor discussed in Sec. III D
suggests that there can be, in principle, 
quantum corrections of the order of 
$e^{2} \sim \epsilon^{3/2}$ at the three-loop order,
which is larger than the corrections at the two-loop order.
However, this does not mean that the expansion is uncontrolled.
If $L$-loop corrections are indeed suppressed only by 
$e^{2/3L} \sim \epsilon^{L/2}$,
one has to compute up to $2n$-loop level in order 
to compute critical exponents to the order of $\epsilon^n$.

\section{Physical Properties}

\subsection{Thermodynamic quantities }

Observables that are local in momentum space,
such as the self energy of a fermion near the Fermi surface
and scattering amplitudes with small momentum exchange,
are insensitive to other modes which are separated in the momentum space.
This is due to the emergent locality in the momentum space\cite{polchinski,LEE2008}, 
which makes the patch description valid in non-Fermi liquid states.
Therefore  temperature dependences of the quantities that are local in momentum space
are solely dictated by their scaling dimensions.

The scaling of thermodynamic quantities are different 
from those observables that are local in momentum space.
This is because all low energy modes near the Fermi surface contribute
to the thermodynamic responses.
In order to examine the scaling behavior of thermodynamic quantities,
we consider the free energy density at finite temperature.
The scaling dimension of the free energy density 
is set by the dimension of  spacetime, $(d-1)z + 1 + 1/2$.
If the free energy was insensitive to all UV cut-off scales,
one would have the form of $f(T) \sim T^{ (d-1) + \frac{3}{2z}}$.
However, this is not the case in theories with Fermi surface
because the free energy is a global quantity 
which depends on all low energy modes around the Fermi surface. 
Since low energy effective theory is local in momentum space,
the singular part of the free energy linearly depends on the size of the Fermi surface\cite{LEE2008}, which then leads to a violation of hyperscaling.
In our local patch description, 
the size of the Fermi surface is set by the largest
momentum $\Lambda$ along the $k_d$ direction.
Because $\Lambda$ has scaling dimension $1/2$,
the free energy density should scale 
as $f(T) = \Lambda T^{ (d-1) + \frac{1}{z}}$.

Let us also consider an external field $h^\alpha$ 
which sources the flavor quantum number given by
\bqa
\rho^\alpha & = & 
\psi_{+,i}^\dagger T^\alpha_{ij} \psi_{+,j}
+ 
\psi_{-,i}^\dagger T^\alpha_{ij} \psi_{-,j}.
\label{rho}
\eqa
Note that $\rho^\alpha$ is the physical flavor quantum number
under which $\psi_{+,i}$ and $\psi_{-,j}$ 
transform in the same manner.
Although all components of $\rho^\alpha$ are conserved in $d=2$,
only parts of them are conserved in $d>2$
due to the absence of the axial flavor symmetry.
In the example of $d=3$ with $N=2$,
this can be easily understood from the fact that
the spin triplet pairing leaves 
only $\sigma^y$ as a conserved flavor
among $T^\alpha = \{ I, \sigma_x, \sigma_y, \sigma_z \}$.
For general $N$, only the flavor density with anti-symmetric $T^\alpha$ 
is related to the conserved charge density,
\bqa
\rho_a^\alpha = j_{A0}^\alpha, \mbox{~~ for $(T^\alpha)^T = -T^\alpha$}.
\eqa
The symmetric flavor density $\rho_s^\alpha$ with $(T^\alpha)^T = T$  is not a conserved density.
Although $\rho_a^\alpha$ is not a density of a conserved current, 
it is related to the $(d-1)$-th component of the  conserved current
\bqa
\rho_s^\alpha = j_{A, d-1}^\alpha,
\eqa
where $j_{A,d-1} = i \bar \Psi_i T^\alpha_{ij} \gamma_{d-1} \Psi$.
Because $\rho^\alpha_a$ and $\rho^\alpha_s$ are parts of different components of the conserved current, the fields that couples to them have different scaling dimensions,
\bqa
[ h_a^\alpha] =  z, ~~ [h_s^\alpha] = 1.
\eqa

From the above considerations, we write the scaling form of the free energy density as
\bqa
f(T,h_s,h_a) = \Lambda T^{ (d-1) + \frac{1}{z} } 
\tilde f \left( h_s T^{-1/z}, h_a T^{-1} \right).
\eqa
This leads to the scaling behavior of the specific heat
and the flavor susceptibility, 
\bqa
c & \sim & T^{(d-2)+\frac{1}{z}}, \label{C} \\
\chi_{ss} & \sim & T^{(d-1)-\frac{1}{z}}, \label{chi++} \\
\chi_{aa} & \sim & T^{(d-3)+\frac{1}{z}}.\label{chi--}  
\eqa
Note  that the flavor susceptibility is anisotropic 
because of the absence of the full flavor symmetry.
Nonetheless, their scaling dimensions 
are completely set by the dynamical critical exponent
because they are parts of the conserved current.
In $d=2$, Eqs. (\ref{C}) and (\ref{chi--}) are consistent 
with the results obtained for 
the specific heat and the susceptibility of conserved spin 
in Ref. \cite{Senthil2008}.
For $d>2$, the low temperature response functions are suppressed by 
a higher powers of temperature because of the suppression 
of density of state with a larger co-dimension of Fermi surface.

\subsection{$2k_F$ scattering}

\begin{figure}
\begin{center}
        \includegraphics[height=4cm,width=14cm]{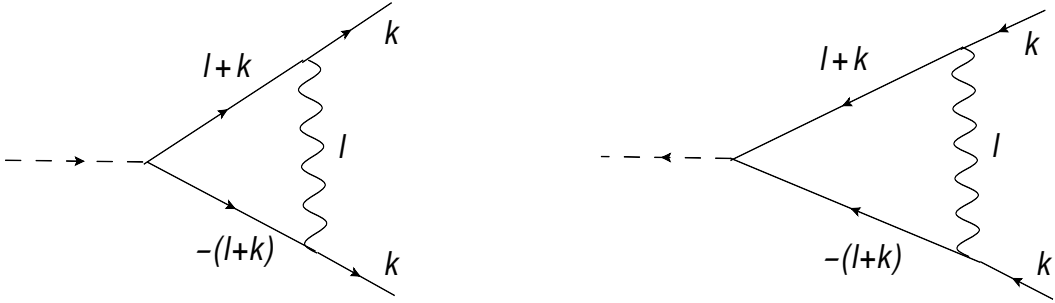} 
\end{center}
\caption{The one-loop diagrams that renormalize the $2k_F$ scattering 
amplitude $r$. 
}\label{fig:scatt2k}
\end{figure}

In order to examine how the back-scattering is affected by interaction 
in the non-Fermi liquid state, we add an operator which carries 
momentum $2 k_F$, 
\bqa\label{scat2k1}
S_{2k_F} & = & -2 \mu r \sum_j \int \frac{dk}{(2\pi)^{d+1}} ( 
(\psi_{+,j}^\dagger (k) \psi_{-,j} (k) + \psi_{-,j}^\dagger (k) \psi_{+,j} (k) ),
\eqa
where $r$ is the source. 
In the spinor representation,
Eq.~(\ref{scat2k1}) can be written as
\bqa
S_{2k_F} & = & i \mu r \int \frac{dk}{(2\pi)^{d+1}} ( 
\Psi^T (k) \gamma_0 \Psi (-k) + \bar{\Psi} (k) \gamma_0 \bar{\Psi}^T (-k) ).
\eqa
To cancel UV divergences, 
we need to add a counter term of the same form,
\bqa
S_{2k_F}^{CT} & = & i \mu r (Z_r - 1) \int \frac{dk}{(2\pi)^{d+1}} 
( \Psi^T (k) \gamma_0 \Psi (-k) + \bar{\Psi} (k) \gamma_0 \bar{\Psi}^T (-k) ),
\eqa
which renormalizes the insertion into
\bqa
S_{2k_F}^{ren} & = & i r_B \int \frac{dk_B}{(2\pi)^{d+1}} ( 
\Psi_B^T (k) \gamma_0 \Psi_B (-k) + \bar{\Psi}_B (k) \gamma_0 \bar{\Psi}_B^T (-k) ),
\eqa
where $r_B = \mu Z_r Z_\psi^{-1}  \left( \frac{Z_2}{Z_1} \right)^{(d-1)} r$
with $Z_r = 1 + Z_{r,1}/\epsilon + ...$. 
The beta function of $r$ is given by
\bqa
\beta_r & = & - ( 1 - \gamma_r ) r,
\eqa
where $\gamma_r  =  \frac{e}{2}z ( Z_{r,1}^{'} - Z_{2,1}^{'} )$ is
the anomalous dimension of the operator. We can easily calculate $Z_{r,1}$ at 
the one-loop level. The diagrams that renormalize $r$ are shown in 
Fig.~\ref{fig:scatt2k}. 
Calculations are done in Appendix~\ref{app:oneloop2kscatt}, 
where it is shown that 
\beq
Z_r = 1+ \frac{e^{4/3}}{N} \frac{u_r}{\epsilon},
\eeq
with the positive value of $u_r$ given by Eq.~(\ref{ur1}).
From this we obtain the anomalous dimension, 
\bqa
\gamma_r = 2 \epsilon + 11.2059 \epsilon^2.
\eqa
As expected, the quantum correction suppresses the $2k_F$ scattering at low energies.

\section{Conclusion}

In summary, we develop a dimensional regularization scheme
where Fermi surface of dimension one 
is embedded in general dimensions
by combining low energy fermionic excitations on opposite patches
of Fermi surface into a Dirac fermion.
When Fermi surface is coupled with a critical boson whose momentum 
is centered at zero, 
the Yugawa coupling becomes marginal at a critical space dimension $d_c=5/2$.
Using $\epsilon = 5/2 - d$ as a perturbative parameter,
we show that the Ising-nematic phase transition
is described by a stable non-Fermi liquid fixed point
near the critical dimension.
Critical exponents and 
temperature dependences of thermodynamic quantities
are computed to the two-loop order.

The dimensional regulairzation scheme is complimentary 
to other expansion schemes\cite{nayak,mross}.
The pro of the dimensional regularization scheme 
is that the locality is maintained in the regularization.
The con is that some symmetry is broken by regularization.
In the expansion scheme based on dynamical modification,
one has to give up some locality in the action, 
but the original symmetry can be easily kept.
Despite the difference in the approach, 
both schemes provide similar conclusions
regarding the existence of stable non-Fermi liquid fixed points
in the perturbative limit and the absence of 
anomalous dimension of boson up to the three-loop order.
In the dimensional regularization scheme, 
there is a room for the boson to acquire 
a non-trivial anomalous dimension 
through a renormalization of the kinetic  term
because all operators in the local action 
can in principle receive quantum corrections
unless protected by a symmetry. 
It is an open question at which order
the anomalous dimension first shows up.

The dimensional regularization scheme 
may be applied to different systems.
However, the direct application of this scheme 
to quantum electrodynamics at finite density
is subtle because of the fact that the superconducting order 
introduced by the dimensional regularization scheme gaps out 
the gauge field.
It will be interesting to find an alternative scheme where 
the global $U(1)$ symmetry is preserved by regularization.
For quantum critical points 
associated with spin/charge density wave\cite{metlsach2},
the critical dimension turns out to be $d_c=3$ 
in the dimensional regularization scheme.
In this case, one does not need to break 
the global $U(1)$ or the spin rotational symmetry 
because one can linearize the dispersion 
of fermions near the hot spots\cite{shouvik2}. 

\section{Acknowledgment}

We would like to thank
Grigory Bednik,
Matthew Fisher,
Patrick Lee,
Sri Raghu,
Subir Sachdev 
and 
T. Senthil
for useful comments and discussions.
The research of SSL was supported in part by 
the Natural Sciences and Engineering Research Council of Canada,
the Early Research Award from the Ontario Ministry of Research and Innovation
and the Templeton Foundation.
Research at the Perimeter Institute is supported 
in part by the Government of Canada 
through Industry Canada, 
and by the Province of Ontario through the
Ministry of Research and Information.

\appendix



\section{Computation of Feynman diagrams at one loop}

\subsection{Boson self-energy}
\label{app:oneloopbos}

Here we compute the one-loop self-energy of boson,
\beq\label{bosloop}
\Pi_1 (q) = -(ie)^2 (d-1) 
\mu^\epsilon 
\int \frac{d^{d+1}k}{(2\pi)^{d+1}} \mbox{Tr}
\left[ \gamma_{d-1} G_0 (k+q)\gamma_{d-1} G_0 (k) \right],
\eeq
where $G_0 (k)$ is the bare fermion propagator given by Eq.~(\ref{propf}). 
Since we are interested in $2\le d < 3$, 
we use the formulas for the $2 \times 2$ gamma matrices,
\bqa
{\rm Tr} \{ \gamma_i \} & = & 0, \nn
{\rm Tr} \{  \gamma_i  \gamma_k  \} & = & 2 \delta_{i k}, \nn
{\rm Tr} 
\{ \gamma_i \gamma_k \gamma_l \gamma_m \} & = &
2( \delta_{ik}\delta_{lm}- \delta_{il}\delta_{km} +
\delta_{im}\delta_{kl} ),
\eqa
where the indices run from $0$ to $d-1$.
The general strategy of computation that applies not only to
Eq. (\ref{bosloop}) but also to all other Feynman diagrams
is to perform the integrations over $k_{d-1}$ and $k_d$
explicitly and then over $\vec K$
in general dimensions.

From the commutation relations between the $\gamma$-matrices, 
we write the self energy as 
\beq\label{pia1}
\Pi_1 (q) = 2 e^2 \mu^\epsilon (d-1) \int \frac{d^{d+1}k}{(2\pi)^{d+1}} 
\frac{\vec K \cdot (\vec K +\vec Q) - \delta_k \delta_{k+q}}
{[\vec K^2 + \delta_k^2]\, [(\vec K +\vec Q)^2 + \delta_{k+q}^2 ]},
\eeq
where
$\delta_k$ and $\delta_{k+q}$ are defined as
\beq
\delta_k = k_{d-1} +\sqrt{d-1} k_d^2, \qquad
\delta_{k+q} = k_{d-1} +q_{d-1} +\sqrt{d-1} (k_d + q_d)^2.
\eeq
It is straightforward to do the integration over $k_{d-1}$ using
the formulas in Eq.~(\ref{useint}) prsented in Appendix~\ref{app:integ},
and obtain 
\beq
\Pi_1 (q) = 2 e^2 \mu^\epsilon (d-1) \int \frac{dk_d d\vec K}{(2\pi)^d} 
\frac{\left( |\vec K +\vec Q|+|\vec K| \right)\,  
\left[ \vec K \cdot (\vec K +\vec Q) - |\vec K|\, |\vec K +\vec Q|\right]}
{ 2 |\vec K|\,|\vec K +\vec Q|\, 
\left[ (\delta_q +2 \sqrt{d-1}  q_d k_d )^2 +(|\vec K +\vec Q|+|\vec K|)^2
\right] }.
\eeq
Making a change of variable, 
\beq
\delta_q +2k_d q_d \sqrt{d-1} \rightarrow {\tilde k}_d,
\eeq
and integrating over ${\tilde k}_d$, 
we arrive at the result, 
\beq\label{pia11}
\Pi_1 (q) = \frac{e^2 \mu^\epsilon 
\sqrt{d-1}}{4|q_d|} I_1 (d-1, \vec Q), 
\eeq
where
\beq
I_1 (d-1, \vec Q) = \int \frac{d\vec K}{(2\pi)^{d-1}} \left\{
\frac{\vec K \cdot (\vec K +\vec Q)}{|\vec K|\,\,|\vec K +\vec Q|}
-1 \right\}. 
\eeq
The $d-1$-dimensional integral in $I_1 (d-1,\vec Q)$ can be done 
using the Feynman parametrization,
\beq\label{feynm}
\frac{1}{A^{\alpha} B^{\beta}}= 
\frac{\Gamma (\alpha +\beta)}{\Gamma (\alpha) \Gamma (\beta)}
\int_0^1 \frac{x^{\alpha-1}\,(1-x)^{\beta-1}\,dx}
{\left[ x A +(1-x) B\right]^{\alpha+\beta}},
\eeq
where $\alpha=\beta=1/2$ and
\beq\label{ABs}
A=\sum_{\mu=0}^{d-2} (k_\mu +q_\mu)^2, \qquad\quad
B=\sum_{\mu=0}^{d-2} k_\mu^2.
\eeq
A change of variables $\vec K \rightarrow \vec K-x \vec Q$
leads to
\beq\label{intI}
I_1 (d-1, \vec Q) = \int \frac{d\vec K}{(2\pi)^{d-1}} \cdot
\frac{1}{\pi} \int_0^1 \frac{dx}{\sqrt{x(1-x)}}\,
\left\{ \frac{\vec K^2 - x (1-x) \vec Q^2}{\vec K^2 + x (1-x) \vec Q^2} 
-1 \right\}.
\eeq
Rescaling $\vec K \rightarrow \sqrt{x (1-x)} \vec K$ and integrating  
over $x$ using the formula
\beq
\int_0^1 \left[ x(1-x)\right]^{(d/2)-1} dx = 
\frac{\Gamma^2 (d/2)}{\Gamma (d)},
\eeq
we obtain 
\beq\label{inti2}
I_1 (d-1, \vec Q) = -\frac{\Gamma^2 (d/2)\, Q^2}
{2^{d-3} \pi^{(d+1)/2} \Gamma (\frac{d-1}{2}) \Gamma (d)}
\int_0^{\infty} \frac{K^{d-2} dK}{K^2 +Q^2}.
\eeq
The integral over $K$ is convergent for $2\le d< 3$ and 
is equal to $-\pi/(2Q^{3-d} \cos\pi d/2 )$.
As a result, the boson self-energy is 
\beq
- \Pi_1 (q) = \beta_d e^2 \mu^\epsilon \frac{|\vec Q|^{d-1}}{|q_d|},
\eeq
with 
\beq
\beta_d = \frac{\sqrt{d-1} \Gamma^2 (d/2)}
{2^d \pi^{(d-1)/2}\, |\cos(\pi d/2)|\, \Gamma(\frac{d-1}{2}) \Gamma (d)}.
\eeq
Note that $\beta_d$ is singular at $d=3$,
which is due to a logarithmic UV divergence
in the coefficient of the Landau damping.
However this is not relevant to us 
because we are concerned about $d$ below $5/2$.

In Eq. (\ref{pia1}), 
one may attempt to perform integrations 
by treating $\delta_k$ and $\delta_{k+q}$ 
as independent variables.
However, this change of variables,
which gives rise to a spurious UV divergence,
is not justified because the integrations 
over $\delta_k$ and $\delta_{k+q}$ 
are not strictly UV convergent,
while the integration over the original 
variables are convergent.

\msk\msk

\subsection{Fermion self-energy}
\label{app:oneloopferm}

Here we compute the one-loop fermion self energy.
From Eqs.~(\ref{fermiloop}) and (\ref{propbos}), the self energy is written as
\beq
\Sigma_1 (q) = i\frac{e^2 \mu^{\epsilon}}{N} (d-1) \int \frac{dk}{(2\pi)^{d+1}} 
D_1 (k) \frac{\gamma_{d-1} \delta_{q-k} -\vec \Gamma \cdot (\vec Q -\vec K)}
{(\vec Q -\vec K)^2 +\delta_{q-k}^2}.
\eeq  
Shifting the variable 
$k_{d-1} \rightarrow k_{d-1}+q_{d-1} +(q_d +k_d)^2$ and integrating over 
$k_{d-1}$ and $k_d$ using  
\beq\label{intovery}
\frac{1}{2\pi}\int_{-\infty}^{\infty} \frac{|x| dx}{|x|^3 +a^2}
=\frac{2}{3\sqrt{3}}\, \frac{1}{a^{2/3}},
\eeq
we obtain 
\beq\label{ferm11}
\Sigma_1 (q) = i\frac{e^{4/3} \mu^{2\epsilon/3}}{N} \frac{(d-1)}
{3\sqrt{3} \beta_{d}^{1/3}} \, I_2 (d-1, \vec Q), 
\eeq
where
\beq
I_2 (d-1, \vec Q) = \int \frac{d\vec K}{(2\pi)^{d-1}}
\frac{\vec \Gamma\cdot (\vec K -\vec Q)}{|\vec K|^{(d-1)/3}\,|\vec K -\vec Q|},
\eeq
and $\beta_{d}$ is given by Eq.~(\ref{betad}).
The $d-1$-dimensional integral in $I_2 (d-1,\vec Q)$ can be calculated using 
the Feynman parametrization (\ref{feynm}). 
For $\alpha=1/2$, $\beta = (d-1)/6$, and $A$, $B$ given by Eq.~(\ref{ABs}),
we obtain 
\beq\label{intII}
I_2 (d-1, \vec Q) = -\frac{\Gamma (\frac{d+2}{6})}
{\Gamma (\frac{d-1}{6}) \sqrt{\pi}} 
\int_0^1 \frac{dx\,(1-x)^{(d-1)/6} }{\sqrt{x}}\,
\int \frac{d\vec K}{(2\pi)^{d-1}} 
\frac{(\vec \Gamma \cdot \vec Q)}{[\vec K^2 + x (1-x) \vec Q^2]^{(d+2)/6}}
\eeq
after a change of variable $k_\mu \rightarrow k_\mu +x q_\mu$.
Rescaling $\vec K \rightarrow \sqrt{x (1-x)} \vec K$ and
integrating over $x$ lead to
\beq
I_2 (d-1, \vec Q) =- 
\frac{\Gamma (\frac{d+2}{6}) \Gamma(\frac{d-1}{3}) \Gamma (\frac{d}{2})}
{\sqrt{\pi} \Gamma (\frac{d-1}{6}) \Gamma(\frac{5d-2}{6})}
\, (\vec \Gamma \cdot \vec Q)\, \int \frac{d\vec K}{(2\pi)^{d-1}} 
\frac{1}{[\vec K^2 + \vec Q^2]^{(d+2)/6}}.
\eeq
From the $(d-1)$-dimensional integration,
\bqa
\int \frac{d\vec K}{(2\pi)^{d-1}} 
\frac{1}{[\vec K^2 + \vec Q^2]^{(d+2)/6}}& =&
\frac{2\pi^{(d-1)/2}}{(2\pi)^{d-1} \Gamma (\frac{d-1}{2})}\,
\int \frac{K^{d-2} \, dK}{[K^2 + Q^2]^{(d+2)/6}} \nonumber\\
&=&
\frac{\Gamma (\frac{5-2d}{6})}{2^{d-1} \pi^{(d-1)/2} \Gamma (\frac{d+2}{6})}
\cdot\frac{1}{Q^{(5-2d)/3}},
\eqa
the self energy is obtained to be
\beq\label{sigma3}
\Sigma_1 (q)  =  - i\frac{e^{4/3}}{N} 
\left( \frac{\mu}{Q}\right)^{2\epsilon/3}\,
\frac{(d-1)\Gamma(\frac{5-2d}{6})\Gamma(\frac{d-1}{3})\Gamma(\frac{d}{2})}
{3\sqrt{3} \beta_{d}^{1/3} \, 2^{d-1} \pi^{d/2}\, \Gamma(\frac{d-1}{6})
\Gamma(\frac{5d-2}{6})} \,
(\vec \Gamma \cdot \vec Q).
\eeq
For small $\epsilon$, $\Gamma(\frac{5-2d}{6})\approx 3/\epsilon$. 
In the $\epsilon\rightarrow 0$ limit, the self energy becomes
\beq\label{sigma4}
\Sigma_1 (q) = 
\left( - \frac{e^{4/3} }{N} \frac{u_1}{\epsilon} 
+ \mbox{ finite terms}
\right)
(i \vec \Gamma \cdot \vec Q),
\eeq
where
\beq\label{u1}
u_1 = \frac{1}
{2^{5/2} \pi^{3/4} \sqrt{3}\, \beta^{1/3}_{5/2} \, \Gamma(\frac34)}.
\eeq

\msk\msk

\subsection{Vertex renormalization}
\label{app:oneloopvert}

At the one-loop level, $A_2$ is zero.
The Ward identity in Eq.~(\ref{WI})
implies that there is no quantum correction to the vertex at the one-loop level.
Here we check this by computing the one-loop vertex correction 
shown in Fig.~\ref{fig:fermvert} (b). 

In general, the fermion-boson vertex function 
$\Gamma_1 (k,q)$ depends on both $k$ and $q$. 
In order to extract the leading $1/\epsilon$ 
divergence, however, it is enough to look at the zero momentum limit,
\beq\label{vertloop}
\Gamma_1 (k,0)= \frac{(ie)^2 \mu^{\epsilon}}{N} (d-1)
\int \frac{dl}{(2\pi)^{d+1}} \gamma_{d-1} G_0 (l)
\gamma_{d-1} G_0 (l)\gamma_{d-1} D_1 (l-k).
\eeq
Using the propagators for fermion and boson in
Eqs.~(\ref{propf}) and (\ref{propbos}), 
and the commutation relation between gamma matrices,
we write the vertex correction as
\beq
\Gamma_1 (k,0)= \frac{e^2 \mu^{\epsilon}}{N} (d-1)
\int \frac{dl}{(2\pi)^{d+1}} D_1 (l-k) \, \gamma_{d-1}\,
\frac{\delta_l^2 - \vec L^2 - 2 \gamma_{d-1} (\vec \Gamma\cdot \vec L) \delta_l}
{[\vec L^2 +\delta_l^2]^2}.
\eeq
One  can readily check that the vertex correction vanishes 
from the identity, $\int_{\infty}^{\infty} (x^2 -a^2)/(x^2 +a^2)^2 =0$.

\section{Computation of Feynman diagrams at two loops}

\subsection{Some useful integrals}
\label{app:integ}

Here we list some integration formulas 
which are useful in the two-loop calculations.

\bqa\label{useint}
\frac{1}{2\pi} \int_{-\infty}^{\infty} 
\frac{(x+a) (x+b) dx}{[(x+a)^2 + A^2][(x+b)^2 + B^2]} =
\frac{|A| |B| (|A|+|B|)}{2 |A| |B| [(a-b)^2 +(|A|+|B|)^2]}, \\
\frac{1}{2\pi} \int_{-\infty}^{\infty} 
\frac{dx}{[(x+a)^2 + A^2][(x+b)^2 + B^2]} =
\frac{(|A|+|B|)}{2 |A| |B| [(a-b)^2 +(|A|+|B|)^2]}, \\
\frac{1}{2\pi} \int_{-\infty}^{\infty} 
\frac{(x+a) dx}{[(x+a)^2 + A^2][(x+b)^2 + B^2]} =
\frac{(a-b) |A|}{2 |A| |B| [(a-b)^2 +(|A|+|B|)^2]}, \\
\frac{1}{2\pi} \int_{-\infty}^{\infty} 
\frac{(x+b) dx}{[(x+a)^2 + A^2][(x+b)^2 + B^2]} =
\frac{(b-a) |B|}{2 |A| |B| [(a-b)^2 +(|A|+|B|)^2]} .
\eqa

\subsection{Boson self-energy}
\label{app:twoloopbos}

Here we compute the two-loop boson self-energy shown 
in Fig.~\ref{fig:bos2} (a),
\bqa\label{prbos2a}
\Pi_2 (q) &=& -\frac{e^4 \mu^{2\epsilon}}{N^2} N (d-1)^2 \,
\int \frac{dl}{(2\pi)^{d+1}} \frac{dp}{(2\pi)^{d+1}} D_1 (l)\nonumber\\
&\times& {\rm Tr} \{ \gamma_{d-1} G_0 (p) \gamma_{d-1} G_0 (p+l) 
\gamma_{d-1} G_0 (p+l+q) \gamma_{d-1} G_0 (p+q) \}.
\eqa
Taking the trace, we obtain
\beq\label{prbos2b}
\Pi_2 (q) = -\frac{e^4 \mu^{2\epsilon}}{N^2} N (d-1)^2 \,
\int \frac{dl}{(2\pi)^{d+1}} \frac{dp}{(2\pi)^{d+1}} D_1 (l)\,
\frac{{\cal B}_{1}}{{\cal D}_{1}},
\eeq
where
\bqa
{\cal B}_{1}& =& 2 [\delta_{p+l} \delta_{p+q+l} -(\vec P +\vec L)
\cdot (\vec P +\vec L +\vec Q)]\, 
[\delta_{p+q} \delta_{p} -(\vec P +\vec Q)
\cdot \vec P] \nonumber\\
&-&2 [(\vec P +\vec L)\cdot (\vec P +\vec Q)]
[(\vec P +\vec L +\vec Q)\cdot \vec P] \nonumber\\
&+&2 [(\vec P +\vec L)\cdot \vec P] 
[(\vec P +\vec L +\vec Q)\cdot (\vec P +\vec Q)] \nonumber\\
&-&2 [\delta_{p+l} (\vec P +\vec L +\vec Q) + \delta_{p+l+q} (\vec P +\vec L)]
\cdot [\delta_{p+q} \vec P + \delta_{p} (\vec P +\vec Q)], \label{B1} \\
{\cal D}_{1} &=& [\delta_{p}^2 +\vec P^2] [\delta_{p+q}^2 +(\vec P +\vec Q)^2]
[\delta_{p+l}^2 +(\vec P +\vec L)^2]
[\delta_{p+l+q}^2 +(\vec P +\vec L +\vec Q)^2]. \label{D1}
\eqa
We then shift the variables as
\beq
p_{d-1} \rightarrow p_{d-1} - \sqrt{d-1} p_d^2, \qquad 
l_{d-1} \rightarrow l_{d-1} - p_{d-1} -\sqrt{d-1} (l_d + p_d)^2,
\eeq
to write
\beq
\delta_{p} \rightarrow p_{d-1}, \qquad \delta_{p+q} \rightarrow p_{d-1}
+2 \sqrt{d-1} p_d q_d +\delta_q,
\eeq
\beq
\delta_{l+p} \rightarrow l_{d-1}, \qquad \delta_{p+l+q} \rightarrow l_{d-1}
+2 \sqrt{d-1} q_d (p_d +l_d) +\delta_q.
\eeq
The integration over $p_{d-1}$, $l_{d-1}$ can be done 
using the formulas given in section~(\ref{app:integ}) 
to obtain
\beq\label{prbos2c}
\Pi_2 (q) = -\frac{e^4 \mu^{2\epsilon}}{N^2} N (d-1)^2 \,
\int \frac{dl_d d\vec L}{(2\pi)^{d}} \frac{dp_d d\vec P}{(2\pi)^{d}} D_1 (l)\,
\frac{{\cal B}_{2}}{{\cal D}_{2}},
\eeq
where
\bqa
{\cal B}_{2} &=& 2 \left( |\vec P +\vec L|+ |\vec P +\vec L +\vec Q| \right)
\left( |\vec P +\vec Q|+ |\vec P| \right)\nonumber\\
&\times& \left. \Bigl\{ 
\left[ |\vec P +\vec L| |\vec P +\vec L +\vec Q|-
(\vec P +\vec L) \cdot (\vec P +\vec L +\vec Q)\right]
\left[ |\vec P +\vec Q| |\vec P|-
(\vec P +\vec Q) \cdot \vec P\right] \right.\nonumber\\
&-& \left. [(\vec P +\vec L)\cdot (\vec P +\vec Q) ]
[(\vec P +\vec L+\vec Q)\cdot \vec P ]
+ [(\vec P +\vec L)\cdot \vec P ]
[(\vec P +\vec L+\vec Q)\cdot (\vec P +\vec Q)] \right. \Bigr\}\nonumber\\
&-&2 \left( 2\sqrt{d-1} q_d (l_d+p_d)+\delta_q \right) 
\left( 2 \sqrt{d-1} p_d q_d +\delta_q \right) \nonumber\\
 &\times&\left[ |\vec P +\vec L +\vec Q | (\vec P +\vec L)- 
|\vec P +\vec L| (\vec P +\vec L +\vec Q )\right]\cdot
\left[ |\vec P +\vec Q | \vec P - 
|\vec P| (\vec P +\vec Q )\right], \label{B2} \\
{\cal D}_{2} &=& 4 |\vec P +\vec L| |\vec P +\vec Q +\vec L| 
|\vec P| |\vec P +\vec Q | \nonumber\\
&\times& \left[ \left( 2 \sqrt{d-1} q_d (l_d +p_d )+\delta_q \right)^2 + 
\left(|\vec P +\vec L|+ |\vec P +\vec Q +\vec L| \right)^2  
\right] \nonumber\\
&\times& \left[ \left( 2 \sqrt{d-1} q_d p_d+\delta_q \right)^2 + 
\left(|\vec P |+ |\vec P +\vec Q | \right)^2  \right]. \label{D2}
\eqa
After we make a further change of variables as
\bqa
\vec L & \rightarrow &  \vec L -\vec P, \nn
\vec P & \rightarrow &  \vec P -\frac{\vec Q}{2}, \nn
2 \sqrt{d-1} q_d p_d  &  \rightarrow &  p_d - \delta_q,
\eqa
and integrate over $p_d$, we obtain 
\beq\label{prbos2d}
\Pi_2 (q) = -\frac{e^4 \mu^{2\epsilon}}{N^2} N (d-1)^2 \,
\int \frac{dl_d d\vec L}{(2\pi)^{d}} \frac{d\vec P}{(2\pi)^{d-1}} 
D_1(l_d,|\vec L -\vec P| )\,
\frac{{\cal B}_{3}(\vec L, \vec P, \vec Q)}
{{\cal D}_{3} (\vec L, \vec P, \vec Q; l_d)},
\eeq
where
\bqa
{\cal B}_{3} (\vec L, \vec P, \vec Q) &=& \Big\{ 
|\vec L -\vec Q/2|+|\vec L +\vec Q/2|+|\vec P -\vec Q/2|
+|\vec P +\vec Q/2| \Big\}\nonumber\\
&\times& \Bigl\{ \left( |\vec L -\vec Q/2||\vec L +\vec Q/2| 
-\vec L^2 +\vec Q^2/4\right)
\left( |\vec P -\vec Q/2||\vec P +\vec Q/2| 
-\vec P^2 +\vec Q^2/4\right) \nonumber\\
&-& \left. \left[(\vec L -\vec Q/2)\cdot (\vec P +\vec Q/2)\right]
\left[(\vec L +\vec Q/2)\cdot (\vec P -\vec Q/2)\right]\right.\nonumber\\ 
&+&\left. \left[(\vec L -\vec Q/2)\cdot (\vec P -\vec Q/2)\right]
\left[(\vec L +\vec Q/2)\cdot (\vec P +\vec Q/2)\right]
\right. \nonumber\\
&-& \left. 
 |\vec L +\vec Q/2| |\vec P +\vec Q/2|
[ (\vec L -\vec Q/2 )\cdot (\vec P -\vec Q/2 )] \right.\nonumber\\
&+&\left. |\vec L +\vec Q/2| |\vec P -\vec Q/2|
[ (\vec L -\vec Q/2 )\cdot (\vec P +\vec Q/2 )] \right.\nonumber\\
&+& \left. |\vec L -\vec Q/2| |\vec P +\vec Q/2|
[ (\vec L +\vec Q/2 )\cdot (\vec P -\vec Q/2 )]\right.\nonumber\\
&-&|\vec L -\vec Q/2| |\vec P -\vec Q/2|
[ (\vec L +\vec Q/2 )\cdot (\vec P +\vec Q/2 )]  \Bigr\}, \label{B3} \\
{\cal D}_{3} (\vec L, \vec P, \vec Q; l_d ) & = &
8 \sqrt{d-1} |q_d| |\vec L-\vec Q/2| |\vec L +\vec Q/2| 
|\vec P-\vec Q/2| |\vec P +\vec Q/2 | \nonumber\\
&& \times \Bigl\{ 4 (d-1) q_d^2 l_d^2 + 
\left(|\vec L-\vec Q/2|+ |\vec L+\vec Q/2| + 
|\vec P-\vec Q/2|+ |\vec P+\vec Q/2|\right)^2  \Bigr\}.  \nn
\label{D3}
\eqa

One can see
that $\Pi_2 (q)$ does not depend on $\delta_q$ 
and vanishes for $\vec Q = 0$. 
It is not difficult to check that 
${\cal B}_{3} =0$ in $d=2$ in agreement with 
Ref.~\cite{metlsach1}, 
although $\Pi_2 (q)$ is non-zero in general dimensions.
To extract the leading behaviour of Eq.~(\ref{prbos2d}) for small $e$, 
we note that the main
contribution to the integral over $l_d$ comes from 
$l_d \sim e^{2/3} |\vec L -\vec P|^{1/2}$, 
which implies that 
$q_d^2 l_d^2 \sim e^{4/3} |\vec L -\vec P| q_d^2$. 
This implies that we can drop the $l_d$ dependence
in ${\cal D}_{3}$ to the leading order in $e$.
Alternatively, one could rescale $l_d \rightarrow e^{2/3} l_d$
and keep the leading order terms in $e$.

In order to extract the dependence on $Q$, 
we write $\vec Q = Q \vec n$, where $\vec n$ is a unit vector,
and rescale the momenta as
\beq
\vec L \rightarrow \vec L \, Q, \qquad \vec P  \rightarrow \vec P \, Q, 
\eeq
to write
\beq\label{prbos2e}
\Pi_2 (q) = -e^2 \mu^\epsilon \frac{Q^{d-1}}{|q_y|} \left[
\frac{e^{4/3}}{N} \left( \frac{\mu}{Q}\right)^{2\epsilon/3} 
\frac{(d-1)^{3/2}}{12 \sqrt3 \beta_d^{1/3}}\,
\int \frac{d\vec L d\vec P}{(2\pi)^{2d-2}}  
\frac{1}{|\vec L -\vec P|^{(d-1)/3}}\,
\frac{{\cal B}_{3}(\vec L, \vec P, \vec n)}
{{\tilde {\cal D}}_{3} (\vec L, \vec P, \vec n; 0)} \right],
\eeq 
to the leading order in $e$, where
\bqa
{\tilde {\cal D}}_{3} (\vec L, \vec P, \vec n; 0) &=&
|\vec L-\vec Q/2| |\vec L +\vec Q/2| 
|\vec P-\vec Q/2| |\vec P +\vec Q/2 | \nonumber\\
&& \times \Bigl\{ 
|\vec L-\vec Q/2|+ |\vec L+\vec Q/2| + 
|\vec P-\vec Q/2|+ |\vec P+\vec Q/2| \Bigr\}^2. 
\eqa
In order to see that Eq.~(\ref{prbos2e}) is UV finite
in $d \le 5/2$, 
let us investigate the behaviour of the integrand for  
$P\gg 1$ and $L\gg 1$. 
Using the fact that
\beq
|\vec L \pm \vec n/2|\approx L \pm \frac{\vec L \cdot \vec n}{2L}
+ \frac{1}{8L} - \frac{(\vec n \cdot \vec L)^2}{8 L^3}, 
\eeq
one obtains 
\bqa
{\cal B}_{3} (\vec L, \vec P, \vec n) &\approx& 
2(L+P) \left\{ (\vec L \cdot \vec P)
-(\vec L \cdot \vec n) (\vec P \cdot \vec n)\right. \nonumber\\
&-& \left.\frac{1}{LP} \left[ 
(\vec L \cdot \vec P)(\vec L \cdot \vec n)(\vec P \cdot \vec n)
+ L^2 P^2 - (\vec L \cdot \vec n)^2 P^2 
-(\vec P \cdot \vec n)^2 L^2 \right] \right\}. 
\eqa
Neglecting the $\vec n$ dependence in ${\tilde {\cal D}}_3$, and using the 
symmetry properties of the integrand under the transformations
$L_\mu \rightarrow -L_\mu$, $P_\mu \rightarrow -P_\mu$, it is easy 
to show that 
\beq
\frac{{\cal B}_{3} (\vec L, \vec P, \vec n)}
{{\tilde {\cal D}}_{3} (\vec L, \vec P, \vec n; 0)} 
\rightarrow \frac{2(L+P)}{(d-1) {\tilde {\cal D}}_{3} (\vec L, \vec P, 0; 0) }
\, \left\{
(d-2) (\vec L \cdot \vec P) -(d-3) LP 
-\frac{(\vec L \cdot \vec P)^2}{LP} \right\}.
\eeq
If we then formally combine $\vec L$ and $\vec P$ into a 
$2(d-1)$-dimensional vector $\vec X =(\vec L, \vec P)$, 
we note that Eq.~(\ref{propbos}) behaves as $\int dX /X^{(17-5d)/3}$ 
at large $X$, which is UV finite.

Now we compute $\Pi_2(q)$ explicitly.
For this, we introduce the $(d-1)$-dimensional spherical coordinate 
in which
the inner products between $\vec n$, $\vec P$, $\vec L$ become
\bqa
\vec P \cdot \vec n &=& P \cos \theta_p, \nn
\vec L \cdot \vec n &=& P \cos \theta_l, \nn
\vec P \cdot \vec L &=& P L ( \cos \theta_p \cos \theta_l + \sin \theta_p \sin \theta_l \cos \phi_l ).
\eqa
In this coordinate system, the integration measure is 
\bqa
d \vec P & = & \frac{ 2 \pi^{\frac{d-2}{2}} }{\Gamma \left( \frac{d-2}{2} \right) } P^{d-2} \sin^{d-3} \theta_p ~dP ~d \theta_p, \nn
d \vec L & = & \frac{ 2 \pi^{\frac{d-3}{2}} }{\Gamma \left( \frac{d-3}{2} \right) } L^{d-2} \sin^{d-3} \theta_l \sin^{d-4} \phi_l ~dL ~d \theta_l ~d \phi_l,
\eqa
and the integration in Eq. (\ref{prbos2e}) becomes 
\bqa\label{prbos2e2}
I^{6a} & = &
\int \frac{d\vec L d\vec P}{(2\pi)^{2d-2}}  
\frac{1}{|\vec L -\vec P|^{(d-1)/3}}\,
\frac{{\cal B}_{3}(\vec L, \vec P, \vec n)}
{{\tilde {\cal D}}_{3} (\vec L, \vec P, \vec n; 0)}  \nn
& = & \int dP dL d \theta_p d \theta_l d \phi_l ~
 \frac{1}{ 2^{2d-4} \pi^{\frac{2d+1}{2}} \Gamma \left( \frac{d-2}{2} \right) \Gamma \left( \frac{d-3}{2} \right) }  \times \nn
&&
\frac{(PL)^{d-2} \sin^{d-3} \theta_p \sin^{d-3} \theta_l \sin^{d-4} \phi_l  }
{ \Bigl( L^2 + P^2 - 2 L P ( \cos \theta_p \cos \theta_l + \sin \theta_p \sin \theta_l \cos \phi_l ) \Bigr)^{(d-1)/6} }
\frac{{\cal B}_{3} (L,P,\theta_p, \theta_l, \phi_l)}{{\tilde {\cal D}}_{3} (L,P,\theta_p, \theta_l)}.  \nn
\eqa 
It is noted that both the measure and the integration over $\phi_l$ 
are ill-defined at $d=5/2$. 
However, these two ill-defined quantities cancel each other in general $d$.
To obtain  a finite result, 
it is important to integrate over $\phi_l$ in general $d$,
and then set $d=5/2$ in the resulting expression.
The rest of the integrations can be done numerically at $d=5/2$, 
which gives
\beq\label{prbos2e3}
\Pi_2 (q) = \beta_d^{6a} e^2  \frac{Q^{d-1}}{|q_y|},
\eq
with
\bq
\beta_d^{6a} \approx 
0.003687 \frac{e^{4/3}}{N}.
\eq

\subsection{Fermion self-energy}
\label{app:twoloopferm}

Here we compute the two-loop contribution to the fermion self-energy 
given by Eq.~(\ref{twoloopf}). 
Simple algebra of the gamma matrices shows that 
the self energy can be divided into two parts,
\beq
\Sigma_2 (q) = \Sigma_{2a} (q) + \Sigma_{2b} (q),
\eeq
where
\bqa
\Sigma_{2a,2b} (q) &=& \frac{i e^4 \mu^{2\epsilon}}{N^2} (d-1)^2 
\int \frac{dp dl}{(2\pi)^{2d+2}} D_1 (p) D_1 (l) \times \nn
&& \frac{{\cal C}_{a,b}}{[(\vec P +\vec Q)^2 +\delta_{p+q}^2]
[(\vec P +\vec L+ \vec Q)^2 +\delta_{p+l+q}^2]
[(\vec L +\vec Q)^2 +\delta_{l+q}^2] }
\eqa
with
\bqa
{\cal C}_{a} & = &\gamma_{d-1} \left\{  \delta_{p+q} \delta_{p+l+q} \delta_{l+q} -
\delta_{l+q} [\vec \Gamma \cdot (\vec P +\vec Q)]
[\vec \Gamma \cdot (\vec P +\vec L+ \vec Q)] \right. \nn
& - & \left.\delta_{p+q} [\vec \Gamma \cdot (\vec P +\vec L+ \vec Q)]
[\vec \Gamma \cdot (\vec L +\vec Q)] - \delta_{p+l+q} 
[\vec \Gamma \cdot (\vec P +\vec Q)] 
[\vec \Gamma \cdot (\vec L +\vec Q)] \right\}, \\
{\cal C}_{b} & = & [\vec \Gamma \cdot (\vec P +\vec Q)] 
[\vec \Gamma \cdot (\vec P +\vec L+ \vec Q)] 
[\vec \Gamma \cdot (\vec L +\vec Q)] \
- \delta_{p+q} \delta_{l+q} [\vec \Gamma \cdot (\vec P +\vec L+ \vec Q)] \nn
& -& \delta_{p+l+q} \delta_{l+q} [\vec \Gamma \cdot (\vec P +\vec Q)]
- \delta_{p+q} \delta_{p+l+q} [\vec \Gamma \cdot (\vec L +\vec Q)].
\eqa
After we shift the variables as
\bqa
p_{d-1} &\rightarrow & p_{d-1} -\delta_q -2\sqrt{d-1} p_d q_d -\sqrt{d-1} p_d^2
\nn
l_{d-1} &\rightarrow & l_{d-1} -\delta_q -2\sqrt{d-1} l_d q_d -\sqrt{d-1} l_d^2,
\eqa
%
we perform the integrations over $p_{d-1}$ and $l_{d-1}$ 
using formulas given in Appendix~\ref{app:integ}
to obtain  
\bqa
\Sigma_{2a} (q) & = & \frac{i e^4 \mu^{2\epsilon}}{N^2} (d-1)^2 
\int \frac{d\vec P d\vec L}{(2\pi)^{2d-2}} \frac{dp_d dl_d}{(2\pi)^{2}} 
D_1 (p) D_1 (l) \times \nn
&& \frac{\gamma_{d-1} (\delta_q -2\sqrt{d-1} p_d l_d)\,\, 
{\bar {\cal C}}_a (\vec L, \vec P, \vec Q)}
{4 \{(\delta_q -2\sqrt{d-1} p_d l_d)^2 +[|\vec P +\vec Q| + |\vec L +\vec Q|
+|\vec P +\vec L +\vec Q| ]^2  \} },  \label{2loosiga} \\
\Sigma_{2b} (q) & = & \frac{i e^4 \mu^{2\epsilon}}{N^2} (d-1)^2 
\int \frac{d\vec P d\vec L}{(2\pi)^{2d-2}} \frac{dp_d dl_d}{(2\pi)^{2}} 
D_1 (p) D_1 (l) \times \nn
&& \frac{[|\vec P +\vec Q| + |\vec L +\vec Q|
+|\vec P +\vec L +\vec Q|]\, \,
{\bar {\cal C}}_b  (\vec L, \vec P, \vec Q)}
{4 \{(\delta_q -2\sqrt{d-1} p_d l_d)^2 +[|\vec P +\vec Q| + |\vec L +\vec Q|
+|\vec P +\vec L +\vec Q| ]^2  \} }, \label{2loosigb}
\eqa
where 
\bqa
{\bar {\cal C}}_a (\vec L, \vec P, \vec Q) & = &
1-\frac{ [\vec \Gamma \cdot (\vec P +\vec Q)] 
[\vec \Gamma \cdot (\vec  P +\vec L +\vec Q)]}{|\vec P +\vec Q|\,
|\vec P +\vec L +\vec Q|} \nn
&-& \frac{ [\vec \Gamma \cdot (\vec P +\vec L +\vec Q)] 
[\vec \Gamma \cdot (\vec L +\vec Q)]}{|\vec P +\vec L + \vec Q|\,
|\vec L +\vec Q|} + 
\frac{ [\vec \Gamma \cdot (\vec P +\vec Q)] 
[\vec \Gamma \cdot (\vec L +\vec Q)]}{|\vec P +\vec Q|\,
|\vec L +\vec Q|}, \\
{\bar {\cal C}}_b (\vec L, \vec P, \vec Q) & = &
\frac{ [\vec \Gamma \cdot (\vec P +\vec Q)] 
[\vec \Gamma \cdot (\vec  P +\vec L +\vec Q)] 
[\vec \Gamma \cdot (\vec L +\vec Q)]}
{|\vec P +\vec Q|\, |\vec P +\vec L +\vec Q| \, |\vec L +\vec Q|} \nn
&-& \frac{[\vec \Gamma \cdot (\vec L +\vec Q)]}{|\vec L +\vec Q|}
+ \frac{[\vec \Gamma \cdot (\vec L + \vec P+ \vec Q)]}
{|\vec L +\vec P+ \vec Q|}
- \frac{[\vec \Gamma \cdot (\vec P +\vec Q)]}{|\vec P +\vec Q|}.
\eqa
We note that that $\Sigma_{2a} (q)$ vanishes for $\delta_q =0$
regardless of the value of $\vec Q$.
On the other hand, $\Sigma_{2a} (q)$ vanishes for $\vec Q=0$. 
Thus we can extract the UV divergent pieces 
by setting $\vec Q =0$ in Eq.~(\ref{2loosiga}) and
expanding the integrand for small $\vec Q$ in Eq.~(\ref{2loosigb}). 
We can also neglect the term $2\sqrt{d-1} p_d l_d$ in the integrands 
to the leading order in $e$ for the same reason discussed 
after Eq. (\ref{D3}).
We then integrate over $l_d$ and $p_d$ to arrive at the following expressions,
\bqa
\Sigma_{2a} (q) &=&  i\gamma_{d-1} \delta_q  \cdot
\frac{e^{8/3} \mu^{4\epsilon/3}}{N^2} \frac{(d-1)^2}{27 \beta_d^{2/3}}
\int \frac{d\vec P d\vec L}{(2\pi)^{2d-2}} 
\frac{{\bar {\cal C}}_a  (\vec L, \vec P, 0)}
{(LP)^{(d-1)/3} \{ \delta_q^2 +[P + L+|\vec P +\vec L| ]^2  \} }, \nn
\label{2loosiga1} \\
\Sigma_{2b} (q) &=&  i (\vec \Gamma \cdot \vec Q)\cdot
\frac{e^{8/3} \mu^{4\epsilon/3}}{N^2} \frac{(d-1)^2}{27 \beta_d^{2/3}}
\int \frac{d\vec P d\vec L}{(2\pi)^{2d-2}} 
\frac{{\cal C}_b^{\prime}  (\vec L, \vec P, \delta_q )}
{(LP)^{(d-1)/3} \{ \delta_q^2 +[P + L+|\vec P +\vec L| ]^2  \} }, \nn
\label{2loosigb1}
\eqa
where
\bqa\label{tcb}
{\cal C}_b^\prime  (\vec L, \vec P, \delta_q ) &=&
\frac{P+L+|\vec P +\vec L|}{PL\,|\vec P +\vec L| }
\left[ \frac{(d-2)}{(d-1)} \left( L^2 +P^2 +(\vec P\cdot \vec L) +PL-
(P+L)|\vec P +\vec L| \right) \right.\nn 
&+& \left.
\frac{2P^2 L^2 -2(\vec P\cdot \vec L)^2}{(d-1)\,|\vec P +\vec L|^2}
\right] + \left[ \frac{\delta_q^2 -[P + L+|\vec P +\vec L| ]^2 } 
{\delta_q^2 +[P + L+|\vec P +\vec L| ]^2 } \right] \times \nn
&&
\left( 1+\frac{(\vec P\cdot \vec L)}{PL} \right)
\frac{(P+L-|\vec P + \vec L|)(P+L+2|\vec P + \vec L|)}
{(d-1)\, |\vec P + \vec L|^2}.
\eqa
In Eq.~(\ref{tcb}),
we use an equality 
$(\vec P \cdot \vec Q) (\vec \Gamma \cdot \vec L) =
(\vec P \cdot \vec L) (\vec \Gamma \cdot \vec Q)/(d-1)$,
which holds inside the integration because
the denominator in Eq. (\ref{2loosigb1}) is 
invariant under the $(d-1)$-dimensional rotation
and the transformations 
$P_\mu \rightarrow -P_\mu$,
$L_\mu \rightarrow -L_\mu$ for each $\mu$.

We then perform the rescaling
\beq
P_\mu \rightarrow P_\mu |\delta_q|, \qquad 
L_\mu \rightarrow L_\mu |\delta_q|,
\eeq 
in Eqs.~(\ref{2loosiga1})-(\ref{tcb}) and introduce the 
spherical coordinate in $(d-1)$ dimensions to
integrate over $d\vec L\, d\vec P$. 
Let $\theta$ be the angle 
between $\vec L$ and $\vec P$. 
Making a change of variables 
\beq
L\rightarrow P l \quad (0<l<\infty), \qquad P\rightarrow P, 
\eeq 
we obtain
\bqa
\Sigma_{2a} (q) &=&  i\gamma_{d-1} \delta_q  \cdot
\frac{e^{8/3}}{N^2} \left( \frac{\mu}{|\delta_q|} \right)^{4\epsilon/3}
\frac{(d-1)^2}{27 \beta_d^{2/3}} \frac{4 \pi^{(2d-3)/2}}{\Gamma (\frac{d-1}{2})
\Gamma (\frac{d-2}{2})}
\int_0^{\infty } \int_0^{\infty } \frac{dl dP}{(2\pi)^{2(d-1)}} \times  \nn
&&
P^{1+2(2d-5)/3} l^{(2d-5)/3} \, \int_0^{\pi}d\theta \, \sin^{d-3}\theta 
\frac{1+\cos\theta}{1+P^2 [1+l+\eta]^2}  \left(
1-\frac{1+l}{\eta} \right), \label{2loosiga2} \\
\Sigma_{2b} (q) &=&  i (\vec \Gamma \cdot \vec Q) \cdot
\frac{e^{8/3}}{N^2} \left( \frac{\mu}{|\delta_q|} \right)^{4\epsilon/3}
\frac{(d-1)}{27 \beta_d^{2/3}} \frac{4 \pi^{(2d-3)/2}}{\Gamma (\frac{d-1}{2})
\Gamma (\frac{d-2}{2})}
\int_0^{\infty } \int_0^{\infty } \frac{dl dP}{(2\pi)^{2(d-1)}} \times \nn
&&
P^{1+2(2d-5)/3} l^{(2d-5)/3} \,
\int_0^{\pi}d\theta \, \sin^{d-3}\theta\, \left\{ 
\frac{1+l+\eta}{1+P^2 (1+l+\eta)^2}  \frac{1}{l\eta} \right. \times  \nn
&& \left.
\left[ (d-2) \left( 1+l^2 + l (1+\cos\theta) -(1+l)\eta \right) +
\frac{2l^2 \sin^2\theta}{\eta^2} \right] \right.\nn 
 &+& \left. \frac{1-P^2 (1+l+\eta)^2}{[1+P^2 (1+l+\eta)^2]^2} \cdot
\frac{(1+l-\eta) (1+\cos\theta) (1+l+2\eta)}{\eta^2} \right\}, \label{2loosigb2}
\eqa
where 
\beq\label{etaa}
\eta \equiv \eta (l,\theta) \equiv \sqrt{1+l^2 + 2l \cos\theta}.
\eeq
In order to extract the leading $1/\epsilon$ contribution  
in Eqs.~(\ref{2loosiga2})-(\ref{2loosigb2}), 
we use  
\beq
\int_0^{\infty} \frac{dP\,P^{1+2(2d-5)/3} }{1+P^2 (1+l+\eta)^2}=
- \frac{\pi}{2\sin\frac{(2d+1)\pi}{3}} 
\frac{1}{(1+l+\eta)^{4(d-1)/3}},
\eeq
\beq
\int_0^{\infty} \frac{dP\,P^{1+2(2d-5)/3} [1-P^2 (1+l+\eta)^2]}
{[1+P^2 (1+l+\eta)^2]^2}=
\frac{(4d-7)\pi}{6\sin\frac{(2d+1)\pi}{3}} 
\frac{1}{(1+l+\eta)^{4(d-1)/3}},
\eeq
and
\beq
\frac{1}{\sin[(2d+1)\pi/3]} \approx - \frac{3}{2\pi \epsilon}.
\eeq
Setting $d=5/2$ everywhere else in the integrands, 
we can single out the UV divergent contributions,
\beq
\Sigma_{2a} (q) = - i \frac{e^{8/3}}{N^2} 
\frac{v_2}{\epsilon} \gamma_{d-1} \delta_q + \mbox{ finite terms} ,
\eeq
\beq
\Sigma_{2b} (q) = - i \frac{e^{8/3}}{N^2} 
\frac{u_2}{\epsilon} (\vec \Gamma \cdot \vec Q)+ \mbox{ finite terms} ,
\eeq
where
\beq\label{v2a}
v_2 = \frac{1}{16\pi^2 \beta_{5/2}^{2/3} \Gamma(3/4) \Gamma (1/4)}
\int_0^{\infty} l\, dl \int_0^\pi \frac{\sin^{3/2}\theta \, d\theta}
{[1+l+\eta]^3 \eta}
\approx 0.000867775,
\eeq
\bqa\label{u2a}
u_2 & = & - \frac{1}{48\pi^2 \beta_{5/2}^{2/3} \Gamma(3/4) \Gamma (1/4)}
\int_0^{\infty} \, dl \int_0^\pi  \frac{\sin^{-1/2}\theta \, d\theta}
{[1+l+\eta]} \left\{ \frac{ 1+l^2 + l (1+\cos\theta) -(1+l)\eta}
{2l\eta} \right.\nn
&& ~~~~~~ +\left. \frac{2l\sin^2\theta}{\eta^3}
- \left( \frac{1+\cos\theta}{\eta^2} \right) \cdot
\frac{(1+l-\eta)(1+l+2\eta)}{1+l+\eta} \right\}
\approx -0.0194218.
\eqa

\section{Computation of the Aslamazov-Larkin-type 
contribution to boson self-energy}
\label{app:ALbos}

The Aslamazov-Larkin-type diagrams shown in Fig.~\ref{fig:ALbos}
give a three loop contribution to boson self-energy,
\beq\label{AL0}
\Pi_{AL} (q) = \Pi_{pp} (q) + \Pi_{ph} (q) = 
\int \frac{dl}{(2\pi)^{d+1}} D_1 (l) D_1 (l-q) f(l,q) 
[f(l,q)+ f(-l,-q) ],
\eeq
where 
\beq\label{f0}
f(l,q) = -\frac{(ie)^3 \mu^{3\epsilon/2}}{N^{3/2}} \, 
N (d-1)^{3/2} \int \frac{dp}{(2\pi)^{d+1}}
{\rm Tr} \{ \gamma_{d-1} G_0 (p+l) \gamma_{d-1} G_0 (p+q) \gamma_{d-1} G_0 (p)\}
\eeq
is the sub-diagram formed by a triangle of a fermion loop.
Since we are interested in quantum correction 
to the local kinetic term of boson, 
we focus on the case of $\vec Q= 0$. 
Taking the trace in Eq.~(\ref{f0}), we obtain
\beq\label{f1}
f(l,q;\vec Q =0) = -\frac{2e^3 \mu^{3\epsilon/2}}{\sqrt{N}} (d-1)^{3/2} 
\int \frac{dp}{(2\pi)^{d+1}} 
\frac{\delta_{p+l}\delta_{p+q}\delta_{p} 
- (\delta_{p}+ \delta_{p+q}) [(\vec P +\vec L)\cdot \vec P]
-\delta_{p+l} P^2 }
{[\delta_{p}^2 +P^2]\, [\delta_{p+q}^2 +P^2]\,
[\delta_{p+l}^2 +(\vec P +\vec L)^2]}.
\eeq
We then make the following shifts of variables
\beq
p_{d-1} + \sqrt{d-1} p_d^2 \rightarrow {\tilde p}_{d-1}, \qquad
2 \sqrt{d-1} p_d q_d +\delta_q \rightarrow {\tilde p}_{d}, 
\eeq
so that 
\begin{gather}
\delta_p \rightarrow {\tilde p}_{d-1}, \qquad
\delta_{p+q} \rightarrow {\tilde p}_{d-1} + {\tilde p}_d \nn
\delta_{p+l} \rightarrow {\tilde p}_{d-1} + \frac{l_d}{q_d} {\tilde p}_d
+\Delta (l, q)
\end{gather}
with $\Delta (l, q) = \delta_l -\frac{l_d}{q_d} \delta_q$.
We then integrate over ${\tilde p}_d$ and ${\tilde p}_{d-1}$,
using a simple generalization of Eqs.~(\ref{useint})
to obtain
\bqa\label{f4}
f(l,q;\vec Q =0) &=& -\frac{e^3 \mu^{3\epsilon/2}}{\sqrt{N}} 
\frac{(d-1)}{q_d} \int \frac{d\vec P}{(2\pi)^{d-1}} \nn
&&\frac{\left( [\vec P \cdot (\vec P +\vec L)] -P\, |\vec P +\vec L|\right) \,
\Delta (l,q) \, [\Theta (l_d) -\Theta (l_d-q_d)]}
{2 P\, |\vec P +\vec L|\, 
\left[ (P+|\vec P +\vec L|)^2 + \Delta^2 (l,q) \right]}.  
\eqa
Note that $f(l,q=0)=0$ follows from Eq.~(\ref{f4}).

For the particle-particle channel which contains $f(l,q) f(l,q)$,
we make a shift 
$l_{d-1} \rightarrow l_{d-1} -\sqrt{d-1} l_d^2 +(l_d/q_d) \delta_q$, and 
integrate over $l_{d-1}$ to obtain
\begin{gather}\label{pipp}
\Pi_{pp}^{} (q) = \frac{e^6 \mu^{3\epsilon}}{N} 
\frac{(d-1)^2}{q_d^2} \int \frac{d\vec P d\vec K}{(2\pi)^{2(d-1)}}
\frac{dl_d d\vec L}{(2\pi)^d} D_1 (l) D_1 (l-q) \times \nn
\frac{\left( [\vec P \cdot (\vec P +\vec L)] -P\, |\vec P +\vec L|
\right)
\left( [\vec K \cdot (\vec K +\vec L)] -K\, |\vec K +\vec L| \right)
[\Theta (l_d) -\Theta (l_d-q_d)]^2}
{8 P K \,|\vec P +\vec L|\,|\vec K +\vec L| 
\left[ P+ |\vec P +\vec L| + K +|\vec K +\vec L| \right]  }.
\end{gather}
To calculate the contribution in the particle-hole channel 
with $f(l,q) f(-l,-q) $ we substitute 
$l_{d-1} \rightarrow l_{d-1} +(l_d/q_d) q_{d-1}$. 
Integration over $l_{d-1}$ gives 
\bqa\label{piph}
\Pi_{ph}^{} (q) &=& -\frac{e^6 \mu^{3\epsilon}}{N} 
\frac{(d-1)^2}{q_d^2} \int \frac{d\vec P d\vec K}{(2\pi)^{2(d-1)}}
\frac{dl_d d\vec L}{(2\pi)^d} D_1 (l) D_1 (l-q) \times \nn
&& \frac{\left( [\vec P \cdot (\vec P +\vec L)] -P\, |\vec P +\vec L|
\right)
\left( [\vec K \cdot (\vec K +\vec L)] -K\, |\vec K +\vec L| \right)
[\Theta (l_d) -\Theta (l_d-q_d)]^2}
{8 P K\,|\vec P +\vec L|\,|\vec K +\vec L|} \times \nn
&& \frac{\left[ P+ |\vec P +\vec L| +K +|\vec K +\vec L| \right]} 
{\left[ P+ |\vec P +\vec L| +K +|\vec K +\vec L| \right]^2 +
4(d-1) l_d^2 (l_d -q_d)^2 }.
\eqa
Note that $\Pi_{pp} (q)$ and $\Pi_{ph} (q)$ 
are  individually UV divergent. 
Their sum, however, leads to a UV finite correction.
Rescaling $l_d$ as
\beq
l_{d} \rightarrow l_d |q_d|
\eeq      
to make the integral over $l_d$ run from $0$ to $1$, and rescaling
\bqa
L_\mu &\rightarrow  & 2 \sqrt{d-1} q_d^2 \, l_d (1-l_d) L_\mu, \nn
P_\mu &\rightarrow  & 2 \sqrt{d-1} q_d^2 \, l_d (1-l_d) P_\mu, \nn
K_\mu &\rightarrow & 2 \sqrt{d-1} q_d^2 \, l_d (1-l_d) K_\mu, 
\eqa
we arrive at the expression in Eq.~(\ref{ALres}).

\section{Renormalization of the $2k_F$ scattering amplitude}
\label{app:oneloop2kscatt}

The diagrams in Fig.~\ref{fig:scatt2k} renormalize
the $2k_F$ scattering amplitude $r$ as 
\beq
\gamma_0 r \rightarrow \gamma_0 r + r \frac{(ie)^2\mu^\epsilon}{N} (d-1) 
\int \frac{dl}{(2\pi)^{d+1}} \gamma_{d-1}^{T} G_0^{T} (l+k)\gamma_0  
G_0 (-l-k) \gamma_{d-1} D_1 (l),
\eeq 
where the superscript $T$ denotes transpose of matrices.
If $d=3$, we have 
$\gamma_0^{T} = -\sigma_y = -\gamma_0$, 
$\gamma_1^{T}= \sigma_z = \gamma_1$ and 
$\gamma_2^T =\sigma_x = \gamma_2$.
For $2<d<5/2$, we generalize this as
\bqa
\gamma_0^{T} &=& -\gamma_0, \nn
\gamma_\mu^T &=& \gamma_\mu,
\mbox{ for  $\mu =1,\ldots,d-1$}. 
\eqa
Using this, we obtain that the one-loop correction,
\beq
\delta r_1 = - r \frac{e^2\mu^\epsilon}{N} (d-1) 
\int \frac{dl_d d\vec L}{(2\pi)^{d}} D_1 (l) 
\frac{(\vec L +\vec K)^2 -\sqrt{d-1} (l_d +k_d)^2 \gamma_{d-1} 
[\vec\Gamma \cdot (\vec L +\vec K )]}{2 |\vec L +\vec K| 
[(d-1)(l_d +k_d)^4 + (\vec L +\vec K)^2]}.
\eeq 
Since one can ignore $l_d$ dependence 
everywhere except for $D_1 (l)$ to the leading order in $e$,
the leading contribution comes from the first term in the
numerator.
For $\vec  K=0$, we obtain
\bqa
\delta r_1 &=& - r \frac{e^{4/3}}{N}\left( \frac{\mu}{\sqrt{d-1} k_d^2} 
\right)^{2\epsilon/3} 
\frac{(d-1)}{3\sqrt{3} \beta_d^{1/3}} \int \frac{d\vec L}{(2\pi)^{d-1}}
\frac{L^{(4-d)/3}}{L^2 +1} \nn
& = & -r \frac{e^{4/3}}{N} \frac{u_r}{\epsilon}
\eqa
with
\beq\label{ur1}
u_r = \frac{\sqrt3}{4\sqrt2 \beta_{5/2}^{1/3} \pi^{3/4} \Gamma (3/4)}
= 0.2627590.
\eeq

\end{document}